\documentclass[ doublespacing]{bmcart}

\usepackage{amsthm,amsmath}
\RequirePackage{natbib}
\RequirePackage{hyperref}
\usepackage[utf8]{inputenc} %unicode support
\usepackage{graphicx}

\startlocaldefs
\endlocaldefs

\begin{document}

%%% Start of article front matter
\begin{frontmatter}

\begin{fmbox}
\dochead{Research}


\title{How does Twitter account moderation work? Dynamics of account creation and suspension on Twitter during major geopolitical events}

%%%%%%%%%%%%%%%%%%%%%%%%%%%%%%%%%%%%%%%%%%%%%%
%%                                          %%
%% Enter the authors here                   %%
%%                                          %%
%% Specify information, if available,       %%
%% in the form:                             %%
%%   <key>={<id1>,<id2>}                    %%
%%   <key>=                                 %%
%% Comment or delete the keys which are     %%
%% not used. Repeat \author command as much %%
%% as required.                             %%
%%                                          %%
%%%%%%%%%%%%%%%%%%%%%%%%%%%%%%%%%%%%%%%%%%%%%%

\author[
   addressref={aff1,aff2},                   % id's of addresses, e.g. {aff1,aff2}                      % id of corresponding address, if any
%   noteref={n1},                        % id's of article notes, if any
   email={francesco.pierri@polimi.it}   % email address
]{\inits{FP}\fnm{Francesco} \snm{Pierri}}
\author[
   addressref={aff1,aff5},
   email={lluceri@isi.edu}
]{\inits{LL}\fnm{Luca} \snm{Luceri}}
\author[
   addressref={aff1,aff3},
   email={echen920@usc.edu}
]{\inits{EC}\fnm{Emily} \snm{Chen}}
\author[
   addressref={aff1,aff3,aff4},
   corref={aff1,aff3,aff4},
   email={emiliofe@usc.edu}
]{\inits{EF}\fnm{Emilio} \snm{Ferrara}}

%%%%%%%%%%%%%%%%%%%%%%%%%%%%%%%%%%%%%%%%%%%%%%
%%                                          %%
%% Enter the authors' addresses here        %%
%%                                          %%
%% Repeat \address commands as much as      %%
%% required.                                %%
%%                                          %%
%%%%%%%%%%%%%%%%%%%%%%%%%%%%%%%%%%%%%%%%%%%%%%

\address[id=aff1]{%                           % unique id
  \orgname{Information Sciences Institute, University of Southern California}, % university, etc
%   \street{Waterloo Road},                     %
  %\postcode{}                                % post or zip code
  \city{Los Angeles},                              % city
  \cny{USA}                                    % country
}
\address[id=aff2]{%
  \orgname{Dipartimento di Elettronica, Informazione e Bioingegneria, Politecnico di Milano},
%   \street{D\''{u}sternbrooker Weg 20},
%   \postcode{24105}
  \city{Milano},
  \cny{Italy}
}
\address[id=aff3]{%                           % unique id
  \orgname{Thomas Lord Department of Computer Science, University of Southern California}, % university, etc
%   \street{Waterloo Road},                     %
  %\postcode{}                                % post or zip code
  \city{Los Angeles},                              % city
  \cny{USA}                                    % country
}
\address[id=aff4]{%                           % unique id
  \orgname{Annenberg School of Communication and Journalism, University of Southern California}, % university, etc
%   \street{Waterloo Road},                     %
  %\postcode{}                                % post or zip code
  \city{Los Angeles},                              % city
  \cny{USA}                                    % country
}
\address[id=aff5]{%                           % unique id
  \orgname{University of Applied Sciences and Arts of Southern Switzerland, Department of Innovative Technologies},
%   \street{},
%   \postcode{}
  \city{Lugano},
  \cny{Switzerland}                         % country
}
%%%%%%%%%%%%%%%%%%%%%%%%%%%%%%%%%%%%%%%%%%%%%%
%%                                          %%
%% Enter short notes here                   %%
%%                                          %%
%% Short notes will be after addresses      %%
%% on first page.                           %%
%%                                          %%
%%%%%%%%%%%%%%%%%%%%%%%%%%%%%%%%%%%%%%%%%%%%%%

% \begin{artnotes}
% %\note{Sample of title note}     % note to the article
% \note[id=n1]{Equal contributor} % note, connected to author
% \end{artnotes}

\end{fmbox}% comment this for two column layout

%%%%%%%%%%%%%%%%%%%%%%%%%%%%%%%%%%%%%%%%%%%%%%
%%                                          %%
%% The Abstract begins here                 %%
%%                                          %%
%% Please refer to the Instructions for     %%
%% authors on http://www.biomedcentral.com  %%
%% and include the section headings         %%
%% accordingly for your article type.       %%
%%                                          %%
%%%%%%%%%%%%%%%%%%%%%%%%%%%%%%%%%%%%%%%%%%%%%%

\begin{abstractbox}

\begin{abstract} % abstract
Social media moderation policies are often at the center of public debate, and their implementation and enactment are sometimes surrounded by a veil of mystery. Unsurprisingly, due to limited platform transparency and data access, relatively little research has been devoted to characterizing moderation dynamics, especially in the context of controversial events and the platform activity associated with them.
Here, we study the dynamics of account creation and suspension on Twitter during two global political events: Russia's invasion of Ukraine and the 2022 French Presidential election. Leveraging a large-scale dataset of 270M tweets shared by 16M users in multiple languages over several months, we identify peaks of suspicious account creation and suspension, and we characterize behaviors that more frequently lead to account suspension. We show how large numbers of accounts get suspended within days of their creation. Suspended accounts tend to mostly interact with legitimate users, as opposed to other suspicious accounts,  making unwarranted and excessive use of reply and mention features, and sharing large amounts of spam and harmful content. While we are only able to speculate about the specific causes leading to a given account suspension, our findings contribute to shedding light on patterns of platform abuse and subsequent moderation during major events.
\end{abstract}

%%%%%%%%%%%%%%%%%%%%%%%%%%%%%%%%%%%%%%%%%%%%%%
%%                                          %%
%% The keywords begin here                  %%
%%                                          %%
%% Put each keyword in separate \kwd{}.     %%
%%                                          %%
%%%%%%%%%%%%%%%%%%%%%%%%%%%%%%%%%%%%%%%%%%%%%%


% MSC classifications codes, if any
%\begin{keyword}[class=AMS]
%\kwd[Primary ]{}
%\kwd{}
%\kwd[; secondary ]{}
%\end{keyword}

\end{abstractbox}
%
%\end{fmbox}% uncomment this for twcolumn layout

\end{frontmatter}

%%%%%%%%%%%%%%%%%%%%%%%%%%%%%%%%%%%%%%%%%%%%%%
%%                                          %%
%% The Main Body begins here                %%
%%                                          %%
%% Please refer to the instructions for     %%
%% authors on:                              %%
%% http://www.biomedcentral.com/info/authors%%
%% and include the section headings         %%
%% accordingly for your article type.       %%
%%                                          %%
%% See the Results and Discussion section   %%
%% for details on how to create sub-sections%%
%%                                          %%
%% use \cite{...} to cite references        %%
%%  \cite{koon} and                         %%
%%  \cite{oreg,khar,zvai,xjon,schn,pond}    %%
%%  \nocite{smith,marg,hunn,advi,koha,mouse}%%
%%                                          %%
%%%%%%%%%%%%%%%%%%%%%%%%%%%%%%%%%%%%%%%%%%%%%%

%%%%%%%%%%%%%%%%%%%%%%%%% start of article main body
% <put your article body there>

%%%%%%%%%%%%%%%%
%% Background %%
%%
\section*{Introduction}
Social media play a major role in modern democracies, enabling individuals to openly discuss political and societal issues as well as respond to crises and emergencies \cite{tang2013facebook, chan2016social}. However, they also expose users to a variety of harmful content that is often promoted by malicious actors in a coordinated fashion \cite{ferrara2016rise, lazer2018science, aral2019protecting, sharma2021identifying,suresh2023tracking}. 
In recent times, we witnessed a rise of hate speech, conspiracy, and disinformation campaigns \cite{ribeiro2018characterizing, bovet2019influence, sharma2022characterizing}. 
An ``infodemic'' of misleading and inaccurate information became particularly worrisome during the COVID-19 pandemic \cite{cinelli2020covid, gallotti2020assessing, yang2021covid, chen2021covid, jiang2021social, rao2021political, chen2022charting,nogara2022disinformation}. Social platforms traditionally take steps to mitigate damage by moderating content and by de-platforming, i.e., removing or suspending accounts that engage in harmful activity \cite{jhaver2021evaluating, chang2022comparative, chen2022charting}. 
Numerous high-profile cases of influential individuals fueling conspiracy theories and inciting violence~\cite{wang2023identifying} have led to real-world incidents and subsequent public outcry \cite{luceri2021social}, bringing platforms to intervene by deactivating accounts of public figures like Donald Trump and Alex Jones.\footnote{theconversation.com/deplatforming-online-extremists-reduces-their-followers-but-theres-a-price-188674}
But moderation has sparked a vivid debate among academics, journalists, and policy-makers, as it might pose threats to freedom of speech \cite{ali2021understanding}. 
A shift toward soft-moderation approaches, which include down-ranking (i.e., lowering the visibility of certain content in other users' feeds), ``shadow banning'' (i.e., not showing that content to other users), and warning labels (i.e., tagging content as potentially harmful or inaccurate), has been recently noticed within platforms' moderation tactics~\cite{papakyriakopoulos2022impact, zannettou2021won}. 

To further our understanding of how these moderation strategies are enacted to tame abuse, here we study Twitter accounts' creation and suspension dynamics. 
Existing research into suspended accounts on Twitter spans various settings, from political elections to social movements~\cite{le2019postmortem,toraman2022blacklivesmatter,majo2021role}: a limitation to these studies is that their analyses are retrospective, i.e., done by querying Twitter's Application Programming Interface (API) with a considerable delay (sometimes years) with respect to the period of activity of observed users, hence preventing the determination of when and why accounts were actually suspended, information that are not disclosed by the platform.
As labeled ground truth data about legitimate and abusive accounts is not always available, researchers typically rely on labeled datasets that emerged during audits or investigations into these platforms \cite{guo2022large}. 
One such example is the case of the Russian \textit{Internet Research Agency} state-controlled accounts \cite{badawy2018analyzing}, whose Twitter handles were released by the US Congress; however, Twitter never disclosed how they identified such malicious actors. This lack of transparency can hinder our understanding of moderation dynamics and perhaps the generalizability of associated findings.
We attempt to address these two problems (retrospective labeling and opaque labels) in this study by timely tracking newly created accounts, and identifying suspended users with minimum delay with respect to the period of observation, with the goal of ascertaining whether and when they were suspended by Twitter. Since monitoring the whole platform would be operationally unfeasible, due to the API's limitations, we focus on two major events that sparked significant attention and are conducive to controversy, e.g., geopolitical events~\cite{stella2018bots, jiang2020political}, hence becoming fertile ground for platform abuse, namely the 2022 Russia-Ukraine war and the 2022 French Presidential election.

The choice of these two events is justified by documented evidence of platform abuse and subsequent Twitter interventions.
For example, after the invasion of Ukraine, researchers and journalists raised attention to a suspicious spike in accounts created on Twitter that were engaging in conversations around the war \cite{osomewp1, osomewp2}. Noticing that many of these accounts were being immediately suspended, it was suggested that the accounts were most likely responsible for the coordinated spread of Russian propaganda. In addition to the geopolitical context provided by the Russo-Ukrainian conflict, we considered a political scenario by focusing on the 2022 French Presidential election, since the previous election in 2017 attracted a considerable amount of malicious activity and coordinated disinformation campaigns \cite{ferrara2017disinformation}.

\subsection*{Research Questions}

Based on these premises, we aim to explore the dynamics of account creation and suspension during the two different major events by collecting information about suspension in a timely fashion. Specifically, we address the following research questions:
\begin{itemize}
    \item \textbf{RQ1}: What are the temporal dynamics of account creation (\textit{RQ1a}) and suspension (\textit{RQ1b}) around major events?
    \item \textbf{RQ2}: Do new and suspended accounts exhibit different behaviours compared to active users (\textit{RQ2a})? What are their patterns of interactions (\textit{RQ2b})?
    \item \textbf{RQ3}: What kind of content do suspended accounts share?
\end{itemize}

We collected and analyzed over 270M tweets in multiple languages, to show that the increase in activity on Twitter during the two major geopolitical events is accompanied by peaks in account creation and abusive behavior, exposing legitimate users to spam campaigns and harmful speech. In particular, we found that Twitter tends to be more proactive towards suspending newly created user accounts, compared to older existing accounts. We also highlight several behavioral features that differentiate users who get suspended from regular and active users, finding very similar results across the two scenarios, and providing insights for research that aims to better understand platforms' policies to handle digital misbehavior and online abuse.

% The outline of the manuscript is the following: we first provide a review of existing literature that relates to this work; then, we describe findings that address the aforementioned research questions; finally, we discuss limitations and draw conclusions.

\section*{Related Work}
In this section, we first review existing contributions focusing on suspended users on Twitter, then we provide an overview of work related to the ongoing conflict in Ukraine and the French Presidential elections.

\subsection*{Suspended Accounts}
Early work on abusive usage of social platforms, which leads to account suspension and removal, mostly focuses on detecting spam, bots, and state-backed trolls \cite{stringhini2010detecting, yang2012analyzing, ferrara2019history, ferrara2022twitter,luceri2020detecting,mazza2022investigating,addawood2019linguistic}. Recently, several contributions carried out retrospective analyses of suspended and deactivated accounts on Twitter in various contexts. 

Le et al. \cite{le2019postmortem} provide a ``postmortem'' analysis of approximately 1M accounts that were active during the 2016 US Presidential election, showing different classes of tweeting behaviors and identifying different communities.

Following a similar approach, Chowdhury et al. \cite{chowdhury2020twitter} identify over 2M suspended accounts that engage mostly in political and marketing campaigns, showing that over 60\% of them were active for more than two years on Twitter before being suspended.
In follow-up work, the same authors \cite{chowdhury2021examining} aim to identify factors that lead to suspension during the 2020 US Presidential election, showing that suspended users use more curse and derogatory words, and tend to share more right-leaning news.

Toraman et al. \cite{toraman2022blacklivesmatter} focus on approximately 500k suspended and deleted users who engaged with the ``Black Lives Matter'' social movement, characterizing their behavior in terms of spam, negative language, hate speech, and misinformation spread.

Seyler et al. \cite{seyler2021textual} tackle the problem of automatically identifying and predicting users' suspension, leveraging tweeting behavior and linguistic cues in the messages shared by suspended and regular users. Leveraging deep neural networks, they achieve up to 82\% accuracy in the binary task of classifying users as suspended or not.

Lastly, Majo et al. \cite{majo2021role} carry out a multi-country analysis of users who got suspended during political elections in 2017 in multiple countries (France, Germany, and the United Kingdom). They show how the behavior and content shared by Twitter suspended accounts are significantly different compared to other active accounts, as they focused more on amplifying divisive issues like immigration and religion.

As we detail next, the present work aims to overcome the main limitation of the above contributions, namely the great amount of delay in the identification and analysis of suspended users with respect to the data collection.  In contrast to previous approaches that studied suspension patterns years after the targeted event, our approach focuses on the analysis of suspended users within weeks. While we explore different dimensions and characteristics of Twitter users compared to the aforementioned analyses, we do report findings that align with such previous results.

\subsection*{2022 Russia-Ukraine War and French Election}
Following the Russian invasion of Ukraine, the Observatory on Social Media at Indiana University investigated the prevalence of suspicious activity on Twitter. In a series of white papers \cite{osomewp1,osomewp2}, they highlight a peak in the creation of new accounts around the day of the invasion, and reveal the presence of coordinated groups of users promoting different campaigns, from boosting the presence of political figures to spam and hate speech. They also show how most of the related messages shared on Twitter are genuine or benign, and that pro-Ukraine messages are much more prevalent than pro-Russia ones.

Caprolu et al. \cite{caprolu2022characterizing} apply mix-methods to analyze a collection of over 5M tweets related to the ongoing conflict, claiming no evidence of massive disinformation campaigns contrary to what was reported in the mainstream news. Park et al. \cite{park2022voynaslov} introduce a dataset called VoynaSlov, which aims to help researchers study Russian language conversations on Twitter and VKontakte (VK), a social platform very popular in Russia, focusing on the attention received by state-affiliated and independent Russian media.

Hanley et al. \cite{hanley2022happenstance} leverage a sentence-level topic analysis technique to study the spread of Russian state propaganda on Reddit, by analyzing the content generated by Russian disinformation websites from January to April 2022. They find approximately 40\% of the comments in the \textit{r/Russia} subreddit are related to Russian disinformation. The same authors \cite{hanley2022special} use a combination of sentiment and topic analysis to study Western, Russian, and Chinese media on Twitter and Weibo, finding that Russian media attempt to justify their ``special military operation'' as opposed to Western press that mostly covers military and humanitarian aspects of the war. Chinese news, instead, insists on the conflict’s diplomatic and economic backlashes.

Geisller et al. \cite{geissler2022russian} study the activity of automated accounts promoting pro- and anti-Russia hashtags on Twitter, in order to quantify the extent to which social bots might influence human accounts by promoting and amplifying Russian propaganda and disinformation.

Pierri et al. \cite{pierri2023propaganda} study the spread of Russian propaganda and misinformation on Facebook and Twitter during the first months of the conflict. They estimate the prevalence of such content on the two platforms, describing temporal patterns and highlighting the disproportionate role played by superspreader accounts. They also estimate the amount of content removed by the two platforms to be around only about 8-15\% of the posts and tweets sharing links to untrustworthy sources.

There is a substantial corpus of literature discussing platform abuse and manipulation on Twitter during political elections in different countries \cite{cinelli2020limited,grinberg2019fake,ferrara2020bots,aral2019protecting,jiang2023retweet}.
For what concerns French Presidential elections, early work by Ferrara \cite{ferrara2017disinformation} studies the presence of disinformation and the role of bots prior to the 2017 election, with a focus on the \textit{MacronLeaks} campaign. They show that users who engaged the most with the campaign were mostly non-French users active in promoting fringe and alt-right narratives, suggesting the possible existence of a black market for reusable political disinformation bots.

Recently, Abdine et al. \cite{abdine2022political} analyze the 2022 French election by showing that supporters of certain candidates engage more in hate speech and aggressive behavior, and revealing the presence of likely bot activity.

\section*{Data Collection}
We employed two different Twitter datasets for our analyses, both collected through the \texttt{Standard v1.1 Streaming} endpoint.\footnote{\url{developer.twitter.com/en/docs/twitter-api/v1}}

The first dataset \cite{chen2022tweets} contains tweets matching keywords related to Russia's invasion of Ukraine -- which occurred on February 24th, 2022 -- in the period from February 22nd to April 28th, 2022. Over 30 keywords in English, Russian and Ukrainian languages were identified by looking at trending topics and hashtags. We refer the interested reader to the related publication \cite{chen2022tweets} for more details on the collection procedure. The data comprises 230,166,962 tweets shared by 14,995,636 unique users. A sample of keywords is available in Table \ref{tab:uk-ru-keywords}, whereas the full list is available in the repository associated with the dataset,\footnote{\url{github.com/echen102/ukraine-russia}} which also contains IDs of tweets that can be re-hydrated querying the Twitter API or using tools like \textit{Hydrator}\footnote{\url{github.com/DocNow/hydrator}} or \textit{twarc}.\footnote{\url{twarc-project.readthedocs.io/en/latest}} We will refer to this dataset as \texttt{UK-RU}. We refer the reader to a similar dataset \cite{munch} which is available at \url{https://github.com/Leibniz-HBI/ukraine_twitter_data}.

The second dataset contains tweets related to the 2022 French Presidential election ---  held on April 10th and April 24th, respectively --- in the period from April 3rd to May 15th 2022. We employed a snowball sampling approach \cite{deverna2021covaxxy} at the end of March to identify 89 relevant keywords in English and French language. The data comprises 39,724,541 tweets shared by 2,792,499 unique users, and it can be accessed in a repository associated with this paper.\footnote{\url{github.com/echen102/fr-elections-2022}} We will refer to this dataset as \texttt{FR-22}.

A limitation in this data collection strategy is the 1\% maximum sampling-rate imposed by Twitter on the streaming endpoint \cite{morstatter2013sample}. The issue of hitting the maximum rate limit occurred occasionally in the case of \texttt{UK-RU}, as it can be seen in the left panel of Figure \ref{fig:tweets-ts}, where the data volume saturates at around 4M daily tweets during the first week of the invasion; it did not arise for \texttt{FR-22} (see the right panel in Figure \ref{fig:tweets-ts}).

We further identified suspended accounts by leveraging the \texttt{POST /2/compliance/jobs} endpoint via \texttt{twarc2}.\footnote{\url{twarc-project.readthedocs.io/en/latest/twarc2\_en\_us/\#compliance-job}} Specifically, on May 23rd we queried Twitter for all the accounts that shared a tweet in \texttt{UK-RU} and \texttt{FR-22}, obtaining almost 2M users that were suspended by the platform for violating their rules. Twitter might suspend an account in a variety of circumstances that range from promoting violence and glorifying crime to hate speech, spam, and impersonation; similarly to other Big Tech platforms, these guidelines are considered among the most stringent \cite{arora2023detecting}.
More details about reasons for suspension are available in the Twitter documentation.\footnote{\url{help.twitter.com/en/managing-your-account/suspended-twitter-accounts}}

\section*{Results}
In the following sections, we provide answers to our research questions from multiple angles. We first look into patterns of account creation and suspension in both datasets. We then analyze the behavioral features and interactions that characterize different classes of accounts. Finally, we describe the type of content shared by suspended users in contrast with legitimate active accounts.

\subsection*{Patterns of Account Creation (RQ1a)}
Panels \textbf{A} and \textbf{B} in Figure \ref{fig:ts-created-suspended} show the daily number of accounts created in \texttt{UK-RU} and \texttt{FR-22}, respectively. We obtained information about account creation from the user object provided by Twitter API, and we are therefore able to count the number of users that were created before the data collection period. In line with \cite{osomewp1, osomewp2}, we notice a peak of accounts created in correspondence with the Russian invasion of Ukraine (February 24th) in \texttt{UK-RU}, followed by a consistent decrease over time. A similar increase is observed in \texttt{FR-22}, and additional peaks are observed in correspondence of the two rounds of elections in \texttt{FR-22}. We also notice a significant peak of account creations in both datasets around April 26th, when Elon Musk announced the deal to buy Twitter. An investigation into this uniquely peculiar peak will be tackled in a separate study.

A total number of 863,017 accounts (5.75\% of all active users in the dataset) were created during the collection period in \texttt{UK-RU}, whereas 80,623 accounts (2.88\% of all active users) were created in \texttt{FR-22}. The proportion in the former dataset is larger, most likely because the conflict captivated a larger audience on a global scale, and because the collection period was slightly longer.

Panels \textbf{C} and \textbf{D} in Figure \ref{fig:ts-created-suspended} show the number of accounts created on each day that were later suspended (as of May 23rd): for \texttt{UK-RU}, 121,548 accounts out of 288,723 suspended accounts ($\sim42.1$\%), whereas for \texttt{FR-22} 7,073 accounts out of 24,805 suspended accounts ($\sim28.5$\%) were created during this period. 

We can notice two different dynamics: in \texttt{UK-RU}, we observe an increase of suspicious accounts created after the invasion, with a few peaks in March followed by a decreasing trend; in \texttt{FR-22} we observe two peaks of creation that precede the two rounds by a few days, which suggest the possible implementation of strategies to pollute online conversations and/or promote specific narratives before the voting events. 
Besides, the proportion of new accounts that get suspended is larger in \texttt{UK-RU} (14\%) than in \texttt{FR-22} (8.77\%). Similarly, the proportion of suspended accounts that are created during the period of observation is much larger in \texttt{UK-RU} (42.1\%) than in \texttt{FR-22} (28.5\%).

One possible explanation could be that the massive audience engaging in online conversations around the conflict consequently lured in a larger amount of malicious activity compared to the French election. These findings confirm previous work which showed peaks of suspensions \cite{le2019postmortem, chowdhury2020twitter} and a larger number of new accounts among those suspended \cite{le2019postmortem,majo2021role,chowdhury2021examining}.

% \textbf{Finding \& Remarks}. Addressing \textit{RQ1a}, we note that a significant fraction of newly-created Twitter accounts gets suspended within the first few weeks from their creation, with this number being much higher for those that engage in chatter pertaining to the geopolitical crisis compared to a national election. A similar result holds when considering the proportion of suspended accounts that are created during the period of observation

\subsection*{Patterns of Account Suspension (RQ1b)}
As Twitter's API does not provide information about the specific time when an account gets suspended, we consider the last appearance of a suspended account as a proxy for suspension time, i.e., the last tweet authored by or targeting (mentioning/replying/retweeting/quoting) the account.

Figure \ref{fig:ts-suspension} shows the daily proportion of accounts that got suspended out of all suspended accounts in each dataset, by labeling accounts created during the collection period as \emph{New}, as opposed to \emph{Old} accounts who were already present on the platform at collection time. 
We can notice slightly different patterns of suspension between the two groups in \texttt{UK-RU} (Pearson $R=0.297$, $p = 0.015$), whereas there is a stronger correlation in \texttt{FR-22} (Pearson $R=0.628$, $p<0.001$). 

Overall, we can notice a few peaks of suspension events for \emph{New} accounts in \texttt{UK-RU} that correspond to peaks in account creation altogether (cf., Figure \ref{fig:ts-created-suspended}), suggesting that a large proportion of those accounts were probably created to deceive, manipulate or pollute online conversations. 
On the other hand, we can observe peaks of suspension in correspondence of the election rounds in \texttt{FR-22}, suggesting a higher level of awareness by the platform with the aim of preserving the integrity of conversations on the election days (cf., the peak of suspension of new accounts a few days before the 2nd round). This result is in line with previous work \cite{le2019postmortem, chowdhury2020twitter}.

We further investigate the lifespan of users who get suspended, both in \textit{relative terms} (i.e., the number of days since their first appearance in the dataset to their suspension), and in \textit{absolute terms} (i.e., the number of days since their creation, for \emph{New} accounts only). 

Panel \textbf{A} in Figure \ref{fig:relative-suspension} shows the distribution of the relative suspension time for both datasets; we observe that \emph{New} accounts get suspended significantly earlier than \emph{Old} ones in both cases. Given that most users shared only a handful of tweets (e.g., 20\% of \emph{New} and \emph{Old} suspended accounts shared only 1 tweet in our datasets), we investigate users' lifespan based on their sharing activity. We perform an exact matching of users based on their number of shared tweets in the two datasets, using the following logarithmic bins: $(1, 10]$, $(10, 100]$, $(100, 1000]$ and $(1000, M]$, where $M$ is the maximum number of tweets shared by a suspended user in each dataset, respectively $M_{\mbox{\texttt{UK-RU}}} = 12,733$ and $M_{\mbox{\texttt{FR-22}}} = 8,002$. We show in Figure \ref{fig:suspended-bins} the number of users present in each bin, for each dataset. We can notice that approximately over 50\% of the accounts shared less than 10 tweets, and that hyper-active users (those with more than 100 tweets shared) are much more prevalent in Old Suspended accounts than in New Suspended accounts, in both datasets.

Panel \textbf{B} in Figure \ref{fig:relative-suspension} shows the distributions of relative lifespan for both classes of suspended users in \texttt{UK-RU}, and similar results are shown in panel \textbf{C} for \texttt{FR-22}; note that we only consider accounts that shared more than 1 tweet. We observe that \emph{New} users get suspended earlier than \emph{Old} users regardless of their tweeting activity, but we also notice that more active users are generally suspended later, accentuating the discrepancy between the two groups. This suggests that the age of an account might be a feature considered by Twitter to promptly detect suspicious accounts at scale, in accordance with previous literature \cite{toraman2022blacklivesmatter,majo2021role,chowdhury2020twitter}.

We observe similar results for new users when considering their absolute lifespan (cf. left panel in Figure \ref{fig:absolute-suspension}, the median lifespan is 1 day in both datasets). In particular, we notice that users in \texttt{FR-22} were suspended earlier than their counterparts in \texttt{UK-RU}, especially those very active (see right panel of the Figure): this might be attributed to the fact that the timescale of the French election event was significantly shorter than that of the Russia-Ukraine conflict and that the event was not so unexpected as the Russian invasion, hence Twitter therein enacted more prompt account suspensions.
% but we argue that this is an artifact likely due the different length of the collection periods. 

We caution that our analysis presents a few caveats, as we only process tweets shared by users in a specific context (dataset) and, thus, we might not be observing the reasons that lead to each given account's suspension. 
Also, given the limited period of analysis, we do not have the same amount of information for each \emph{New} account, i.e., we collect fewer observations for those created toward the end of the collection period compared to those created at the beginning, and the length of the collection period differs for the two datasets.

% \textbf{Finding \& Remarks}. As for \textit{RQ1b}, we established that Twitter suspensions occur mostly within the first few days in the lifespan of newly-created accounts, and affect not only hyper-active accounts, which paradoxically tend to be suspended later than accounts with limited observable activity.

\subsection*{Behaviour of Suspended Accounts (RQ2a)}
% boxplots of features
% inter-tweeting time

We first defined two additional classes to investigate whether suspended users exhibit a different behaviour compared to active users -- \emph{New} and \emph{Old} active accounts -- depending on their creation date. These classes are highly imbalanced as most users belong to the \emph{Old Active} class (93\% in \texttt{UK-RU} and 96\% in \texttt{FR-22}). 
For each group of users, we computed the following sets of features:
\begin{itemize}
    \item Proportion of original tweets, replies, retweets, and quotes out of all their shared tweets.
    \item Initial and final number of followers.
    \item Proportion of \emph{contextual} tweets, i.e., the ratio between the number of tweets shared in the dataset and the final statuses count in their profile, i.e., the total number of tweets shared by the user on Twitter.
    \item Mean inter-tweeting time, i.e., the average time (in seconds) between two consecutive tweets shared by a user.
\end{itemize}
The last feature can be seen as a lower bound of the relative activity of a user in a particular context (dataset) -- despite known limitations derived from sampling biases and deletion activity \cite{torres2022manipulating}.

We first compared \emph{Suspended} versus \emph{Active} users, by looking at the distributions of the aforementioned features with an exact matching procedure based on the number of tweets. We observe in both datasets that \emph{Suspended} accounts made a significantly larger usage of replies compared to \emph{Active} users in both datasets (see Figure \ref{fig:boxplot-replies}). 
Accordingly, \emph{Suspended} accounts retweeted less and shared fewer original tweets/quotes w.r.t. \emph{Active} users. This difference holds also when considering \emph{Old} or \emph{New} users separately, and in both datasets.

We then looked at the differences between \emph{New Suspended} and \emph{New Active} users, and we observe that the latter group exhibits a larger growth in followers (i.e., the final minus initial number of followers) in both datasets but more visible in \texttt{UK-RU}, also when controlling for lifespan (see Figure \ref{fig:growth-followers}).

We also considered the proportion of contextual tweets, which is similar among the two classes, and observe that it increases with the tweeting activity of users, although the median value stays below 50\% in both datasets (see Figure \ref{fig:contextual-tweets-boxplots}). This is because accounts partake to some extent in other discussion topics not captured by our data collection.

% This suggests that we are likely missing a portion of users' activity, which is not captured in our data collection, and it indicates that users likely take part in conversations on multiple topics. This confounding factor might hinder the possibility of detecting reasons for suspension, and addressing it would require collecting the timeline of each user in a continuous fashion.

Lastly, we analyzed the inter-tweeting time of different classes of users, considering only users with at least 10 tweets. As shown in Figures \ref{fig:tweeting-activity-uk} and \ref{fig:tweeting-activity-fr}, we first observe that \textit{Suspended} accounts tweet with significantly higher frequency compared to \emph{Active} users (panel A).
Also, \emph{New} users in general exhibit a much higher level of hyper-activity compared to other classes (panels B and C). Similar considerations hold for both datasets, and are in accordance with the past literature \cite{majo2021role,seyler2021textual}.

% \textbf{Finding \& Remarks}. In response to \textit{RQ2a}, we find that \textit{Suspended} accounts exhibit significantly larger usage of replies, while retweeting less and sharing fewer original tweets; this may suggest that abusive behaviors leading to suspension might occur preferentially by replies. \textit{Suspended} accounts also exhibit unreasonably high frequency of activity.

\subsection*{Patterns of Interactions  of Suspended Accounts (RQ2b)}
Next, we analyze the interaction patterns enacted by \emph{Suspended} and \emph{Active} accounts, by considering replies, retweets, and quotes. Given class imbalance, most of the interactions specifically involve \emph{Active} users, namely over 190 millions and 36 millions, respectively for \texttt{UK-RU} and \texttt{FR-22}. Conversely, just hundreds of thousands of interactions involve solely \emph{Suspended} accounts. We thus normalize the number of interactions by source and target before searching for patterns, i.e., we respectively divide the number of interactions between two groups by the total amount of interactions generated (source) or received (target). Moreover, we employ a null model to statistically assess whether the amount of interactions taking place between different classes of users is larger or smaller than expected in a random ensemble obtained by assigning to users a random class label 100 times.

We show results for \texttt{UK-RU} in Figure \ref{fig:interactions-heatmap-uk-ru}, where we provide a heatmap with statistically significant values. We can see that the amount of interactions originating from \textit{Old Active} users is smaller than expected -- both when normalized by source or target -- whereas interactions between other classes of users are higher than expected. Figure \ref{fig:interactions-heatmap-fr} shows different results for \texttt{FR-22}: interactions normalized by source are mostly not significant, with a larger amount of interactions within \textit{New Suspended} users than expected; interactions normalized by the target are smaller than expected when originating from \textit{Old Active} users, and larger than expected for \textit{New Active}, \textit{Old Suspended} and \textit{New Suspended} users when they interact with users from the same class.

We further looked at the interactions between groups over time. The time series of interactions from/to \emph{Active} users simply reflects the trend shown in Figure \ref{fig:tweets-ts}, in both datasets.
We report a decreasing trend in the number of interactions from/to \emph{Old Suspended} accounts over time in both datasets -- with no specific patterns among different actions -- most likely due to the fact that many were suspended during the period of observation, thus reducing the sample over time and consequently the number of interactions. 

Focusing on interactions taking place among \emph{New Suspended} accounts, as shown in Figure \ref{fig:ts-interactions}, we notice some spikes in the number of replies (sent by these accounts) in both datasets, coherently with the behavior highlighted in the previous section; interestingly, these spikes are aligned with the peak of account creation (cf., Figure \ref{fig:ts-created-suspended}). 
Additionally, we observe spikes of interactions among \emph{New Suspended} accounts on specific days, in both datasets, which might indicate spam activity or other malicious behaviors in need of further investigation.

% \textbf{Finding \& Remarks}. In addressing \textit{RQ2b}, we revealed that suspended accounts mostly interact with, and receive attention from legitimate users. Spikes of out-group interactions align well with pivotal events, revealing a possible mechanism that triggers Twitter's suspension. 

% \begin{figure}[!ht]
%     \centering
%     \includegraphics[width=\linewidth]{img/ts_interactions_new_new.pdf}
%     \caption{Number of interactions between New Suspended users in \texttt{UK-RU} \textbf{(A)} and \texttt{FR-22} \textbf{(B)}. Vertical lines indicate the invasion of Ukraine (black) and the two rounds of elections (green, light orange)}
%     \label{fig:ts-interactions-suspended}
% \end{figure}

\subsection*{Content Characterization of Suspended Accounts (RQ3)}
We analyzed the content shared by \emph{Suspended} and \emph{Active} accounts by first extracting the Uniform Resource Locators (URL) and hashtags most shared during the period of analysis by each of the two macro-groups. We do not report relevant differences when looking at top web domains. For what concerns hashtags: In \texttt{FR-22}, we observe a slightly higher presence of inflammatory hashtags against Macron (e.g. \texttt{ \#toutsaufmacron}) and \texttt{\#touscontremacron}) shared by \emph{Suspended} accounts compared to \emph{Active} ones (see Figure \ref{fig:top-hashtags-fr}, in line with previous research~\cite{ferrara2017disinformation, luceri2019red, stella2018bots};
In \texttt{UK-RU}, we observe that Suspended users shared several hashtags related to Non-Fungible-Tokens (NFT) and cryptocurrency-related spam (e.g., \texttt{\#babydoge} and \texttt{\#shibainu}) see Figure \ref{fig:top-hashtags-uk-ru}), in line with other recent studies \cite{ye2023online,nizzoli2020charting, nghiem2021detecting}. 

% Next, we employed Google Jigsaw Perspective API\footnote{\url{www.perspectiveapi.com/}} to estimate the probability of tweets shared by users to be toxic. In addition to English language tweets, we considered Russian for \texttt{UK-RU}\footnote{Ukrainian is not supported by the API.} and French for \texttt{FR-22}...

Next, two annotators independently (and manually) labeled a random sample of 100 original tweets, replies, or quotes (we intentionally excluded retweets) shared by accounts in each class, namely \emph{New Active}, \emph{Old Active}, \emph{New Suspended} and \emph{Old Suspended} accounts. Annotators did not have information about the class of the user that shared each tweet (i.e., if they were active/suspended or new/old), and there were only 73 coding disagreements out of 800 tweets, which were discussed in order to reach an agreement on a single label. We referred to the labeling taxonomy introduced in \cite{founta2018large}, where authors considered the following categories: Offensive language, Abusive language, Hate speech, Aggressive behavior, Cyberbullying behavior, Spam, and Normal. We collapsed the first 5 categories into one macro-category (Harmful) and thus considered three classes for understanding the type of messages posted by suspended accounts: Harmful, Normal, and Spam. 

We show the resulting proportion of tweets shared by different classes of users for each category (and dataset) in Figure \ref{fig:labeled_tweets}. We find similar results in both datasets: suspended users shared a larger amount of Harmful and Spam messages compared to active users; \textit{New Suspended} users were particularly active in spamming campaigns in both datasets, and these findings are in line with previous work \cite{toraman2022blacklivesmatter,chowdhury2020twitter,chowdhury2021examining}.

% \textbf{Finding \& Remarks}. Answering \textit{RQ3}

%%%%%%%%%%%%%%%%%%%%%%%%%%%%%%%%%%%%%%%%%%%%%%
%%                                          %%
%% Backmatter begins here                   %%
%%                                          %%
%%%%%%%%%%%%%%%%%%%%%%%%%%%%%%%%%%%%%%%%%%%%%%

\section*{Discussion}
\subsection*{Contributions}
We studied the dynamics of account creation and suspension on Twitter during major geopolitical events such as the Russian-Ukraine conflict and the 2022 French Presidential election, showing results that are generalizable to different settings. 
Leveraging a large-scale dataset of 270M tweets in multiple languages shared by over 16M users, we uncovered peaks in the creation of accounts which in some cases were associated with specific events such as the election rounds. These included a large number of users that were later suspended for violating Twitter’s policies. 
We highlighted how Twitter tends to be more proactive towards recently created accounts compared to users with a longer lifespan. 
We analyzed in detail the behavioral features which characterize suspended accounts from legitimate active ones, showing an excessive usage of replies and toxic language by the former group as well as a higher level of activity. 
We studied the interactions between different classes of users, finding that abusive users were only successful at reciprocating with new legitimate ones throughout the period of observation, but not with the old accounts. 
Finally, by means of a qualitative analysis of a small sample of tweets, we estimated that suspended accounts frequently shared harmful and spam messages, which most likely lead to their deactivation.

\subsection*{Limitations}
There are a number of limitations in our study. Due to the 1\% limit in Twitter's Filter API \cite{morstatter2013sample}, we were occasionally unable to capture the full volume of conversations related to the war in Ukraine, potentially missing peaks of account creation and activity; we did not incur in the same issue with the French Presidential election data collection. However, finding many similar results across the two datasets suggests that our data sample was not severely affected by this limitation.
Our list of keywords might not include all the different spellings and transliterations of Ukrainian and Russian words, and our dataset might miss some relevant Twitter activity.
As Twitter does not release details on the reasons behind the suspension nor the timestamp of the event, our proxy approach to detect suspension might be prone to error. It might also be the case that different moderation efforts have been applied in the two settings, and that policies are likely to change over time.
We did not filter out automated accounts, i.e., social bots~\cite{yang2022botometer}, 
and an investigation to relate bot behavior to suspension
% of the correlation between likely-bot behavior and suspension 
is left for future research, considering that a correlation between account suspension and increased bot likeness was already observed in other social and political discussions \cite{ferrara2020types, luceri2021down}.  
Lastly, there are two sources of confounding/unobserved factors when studying the characteristics of users who get suspended. One comes from the scope of our analysis, as we only analyze tweets related to a given topic of conversation, e.g., the ongoing conflict or the election, and users might be engaging in other conversations that are not captured by our data; this would require to collect data from each user in a separate dedicated stream. The other source of confounding factors comes from the fact that many active users might exhibit malicious behaviors similar to suspended users, but have not been flagged or detected yet by the platform.

\subsection*{Conclusions and Future Work}
%% conclusions
% increase of traffic around major events which expose individuals to manipulation and abuse
% need to intervene timely
% need more exploration
Our results show that social media platforms are particularly exposed to digital harm and online manipulation during events that captivate the public discourse on a large scale, when the volume of conversations and user engagement increases rapidly. Our work contributes to the extant literature on the behavioral dynamics of users who pollute social media platforms, but numerous questions still remain open in relation to the behavior of malicious actors and their influence on online conversations. This is particularly relevant given existing limitations to accessing data from social platforms, which do not transparently disclose how they detect and remove harmful content and accounts. 

On the one hand, the spread of false information, hate speech, and other shenanigans on online social media are detrimental to the democratic process. 
On the other hand, platform interventions might be perceived as posing threats to freedom of speech and, as one unintended consequence, deplatforming can cause migration to other fringe communities, which are harder to map and study, leading to an increase in extremism of online activity. Our results call for more transparent collaborations between academics, platforms, and regulators in order to devise effective strategies to cope with online harm and manipulation in a timely manner, especially during major events that involve massive audiences.

%% future work
% - prediction/detection tool
% - coordinated activity framework, understand if there are promoting campaigns toward specific users
% - deeper investigation of the content
% - apply to other datasets
Future work might build upon our findings to design algorithms to automatically predict users that will be violating platforms' terms, as well as spotlighting users that behave maliciously but have not been suspended yet. 
Researchers could further investigate the presence of automated and (inauthentic) coordinated behavior~\cite{sharma2021identifying,suresh2023tracking}, which might involve users that get suspended, and they could perform a deeper investigation of the content shared by these users at scale. Finally, future research could consider multiple platforms simultaneously, and study the role of suspended users in other domains, such as medical or scientific topics. 

\subsection*{Ethical considerations}
In the spirit of transparency and open research, we provide public access to the two datasets collected for this study.
Both datasets consist of public posts collected via APIs that are accessible to the general public. To abide by Twitter's terms of service we only release IDs of tweets. These can be used to retrieve the data analyzed in this paper, with the exception of posts that have been removed or made private by users, thus limiting reproducibility analyses. At the time of this writing, we also acknowledge that access for researchers to Twitter's API might be limited in light of the new policies, and this might hinder future usage of our resources. Following standard ethical guidelines, we did not attempt to identify or de-anonymize users, and we only report aggregate analyses. We acknowledge that malicious actors might exploit our results to better understand platforms' moderation policies and devise strategies to avoid being detected and suspended, especially when conducting harmful campaigns during relevant global events.

\smallskip Note: this project was approved by our institution's IRB.

\begin{backmatter}

\section*{Declarations}

% \subsection*{Acknowledgments}
% The authors are thankful to Emily Chen for providing access to the data.
\subsection*{Abbreviations}
API = Application Programming Interface\\
URL = Uniform Resource Locator

\subsection*{Author's contributions}
FP, LL, and EF conceptualized the research. EC collected the data and contributed to the revision of the manuscript. FP and LL designed the experiments and analysed the data. FP performed research analyses and wrote the initial version of the manuscript. LL and EF supervised the research. All authors wrote, read, and approved the final manuscript.

\subsection*{Availability of data and materials}
We provide full access to IDs of tweets analyzed in our work, which can be retrieved using Twitter's API. The datasets are available \url{github.com/echen102/ukraine-r} and \url{github.com/echen102/fr-elections-2022}.

\subsection*{Competing interests}
The authors declare that they have no competing interests.

\subsection*{Funding}
Work supported in part by DARPA (\#HR001121C0169), PRIN grant HOPE (FP6, Italian Ministry of Education), Swiss National Science Foundation (grant CRSII5\_209250) via the SINERGIA project CARISMA (carisma-project.org/), and the EU H2020 research and innovation programme, COVID-19 call, under grant agreement No. 101016233 “PERISCOPE” (periscopeproject.eu/). Any opinions, findings, and conclusions or recommendations expressed in this paper are those of the authors and do not necessarily reflect the views of the funding agencies.

%%  After submission of the .TEX file,                     %%
%%  you will be prompted to submit your .BBL file.         %%

% if your bibliography is in bibtex format, use those commands:
\bibliographystyle{bmc-mathphys} % Style BST file (bmc-mathphys, vancouver, spbasic).
\bibliography{bib.bib}      % Bibliography file (usually '*.bib' )


\begin{thebibliography}{70}
\ifx \bisbn   \undefined \def \bisbn  #1{ISBN #1}\fi
\ifx \binits  \undefined \def \binits#1{#1}\fi
\ifx \bauthor  \undefined \def \bauthor#1{#1}\fi
\ifx \batitle  \undefined \def \batitle#1{#1}\fi
\ifx \bjtitle  \undefined \def \bjtitle#1{#1}\fi
\ifx \bvolume  \undefined \def \bvolume#1{\textbf{#1}}\fi
\ifx \byear  \undefined \def \byear#1{#1}\fi
\ifx \bissue  \undefined \def \bissue#1{#1}\fi
\ifx \bfpage  \undefined \def \bfpage#1{#1}\fi
\ifx \blpage  \undefined \def \blpage #1{#1}\fi
\ifx \burl  \undefined \def \burl#1{\textsf{#1}}\fi
\ifx \doiurl  \undefined \def \doiurl#1{\textsf{#1}}\fi
\ifx \betal  \undefined \def \betal{\textit{et al.}}\fi
\ifx \binstitute  \undefined \def \binstitute#1{#1}\fi
\ifx \binstitutionaled  \undefined \def \binstitutionaled#1{#1}\fi
\ifx \bctitle  \undefined \def \bctitle#1{#1}\fi
\ifx \beditor  \undefined \def \beditor#1{#1}\fi
\ifx \bpublisher  \undefined \def \bpublisher#1{#1}\fi
\ifx \bbtitle  \undefined \def \bbtitle#1{#1}\fi
\ifx \bedition  \undefined \def \bedition#1{#1}\fi
\ifx \bseriesno  \undefined \def \bseriesno#1{#1}\fi
\ifx \blocation  \undefined \def \blocation#1{#1}\fi
\ifx \bsertitle  \undefined \def \bsertitle#1{#1}\fi
\ifx \bsnm \undefined \def \bsnm#1{#1}\fi
\ifx \bsuffix \undefined \def \bsuffix#1{#1}\fi
\ifx \bparticle \undefined \def \bparticle#1{#1}\fi
\ifx \barticle \undefined \def \barticle#1{#1}\fi
\ifx \bconfdate \undefined \def \bconfdate #1{#1}\fi
\ifx \botherref \undefined \def \botherref #1{#1}\fi
\ifx \url \undefined \def \url#1{\textsf{#1}}\fi
\ifx \bchapter \undefined \def \bchapter#1{#1}\fi
\ifx \bbook \undefined \def \bbook#1{#1}\fi
\ifx \bcomment \undefined \def \bcomment#1{#1}\fi
\ifx \oauthor \undefined \def \oauthor#1{#1}\fi
\ifx \citeauthoryear \undefined \def \citeauthoryear#1{#1}\fi
\ifx \endbibitem  \undefined \def \endbibitem {}\fi
\ifx \bconflocation  \undefined \def \bconflocation#1{#1}\fi
\ifx \arxivurl  \undefined \def \arxivurl#1{\textsf{#1}}\fi
\csname PreBibitemsHook\endcsname

%%% 1
\bibitem{tang2013facebook}
\begin{barticle}
\bauthor{\bsnm{Tang}, \binits{G.}},
\bauthor{\bsnm{Lee}, \binits{F.L.}}:
\batitle{Facebook use and political participation: The impact of exposure to
  shared political information, connections with public political actors, and
  network structural heterogeneity}.
\bjtitle{Social science computer review}
\bvolume{31}(\bissue{6}),
\bfpage{763}--\blpage{773}
(\byear{2013})
\end{barticle}
\endbibitem

%%% 2
\bibitem{chan2016social}
\begin{barticle}
\bauthor{\bsnm{Chan}, \binits{M.}}:
\batitle{Social network sites and political engagement: Exploring the impact of
  facebook connections and uses on political protest and participation}.
\bjtitle{Mass communication and society}
\bvolume{19}(\bissue{4}),
\bfpage{430}--\blpage{451}
(\byear{2016})
\end{barticle}
\endbibitem

%%% 3
\bibitem{ferrara2016rise}
\begin{barticle}
\bauthor{\bsnm{Ferrara}, \binits{E.}},
\bauthor{\bsnm{Varol}, \binits{O.}},
\bauthor{\bsnm{Davis}, \binits{C.}},
\bauthor{\bsnm{Menczer}, \binits{F.}},
\bauthor{\bsnm{Flammini}, \binits{A.}}:
\batitle{The rise of social bots}.
\bjtitle{Communications of the ACM}
\bvolume{59}(\bissue{7}),
\bfpage{96}--\blpage{104}
(\byear{2016})
\end{barticle}
\endbibitem

%%% 4
\bibitem{lazer2018science}
\begin{barticle}
\bauthor{\bsnm{Lazer}, \binits{D.M.}},
\bauthor{\bsnm{Baum}, \binits{M.A.}},
\bauthor{\bsnm{Benkler}, \binits{Y.}},
\bauthor{\bsnm{Berinsky}, \binits{A.J.}},
\bauthor{\bsnm{Greenhill}, \binits{K.M.}},
\bauthor{\bsnm{Menczer}, \binits{F.}},
\bauthor{\bsnm{Metzger}, \binits{M.J.}},
\bauthor{\bsnm{Nyhan}, \binits{B.}},
\bauthor{\bsnm{Pennycook}, \binits{G.}},
\bauthor{\bsnm{Rothschild}, \binits{D.}}, \betal:
\batitle{The science of fake news}.
\bjtitle{Science}
\bvolume{359}(\bissue{6380}),
\bfpage{1094}--\blpage{1096}
(\byear{2018})
\end{barticle}
\endbibitem

%%% 5
\bibitem{aral2019protecting}
\begin{barticle}
\bauthor{\bsnm{Aral}, \binits{S.}},
\bauthor{\bsnm{Eckles}, \binits{D.}}:
\batitle{Protecting elections from social media manipulation}.
\bjtitle{Science}
\bvolume{365}(\bissue{6456}),
\bfpage{858}--\blpage{861}
(\byear{2019})
\end{barticle}
\endbibitem

%%% 6
\bibitem{sharma2021identifying}
\begin{bchapter}
\bauthor{\bsnm{Sharma}, \binits{K.}},
\bauthor{\bsnm{Zhang}, \binits{Y.}},
\bauthor{\bsnm{Ferrara}, \binits{E.}},
\bauthor{\bsnm{Liu}, \binits{Y.}}:
\bctitle{Identifying coordinated accounts on social media through hidden
  influence and group behaviours}.
In: \bbtitle{KDD’21}
(\byear{2021})
\end{bchapter}
\endbibitem

%%% 7
\bibitem{suresh2023tracking}
\begin{botherref}
\oauthor{\bsnm{Suresh}, \binits{V.P.}},
\oauthor{\bsnm{Nogara}, \binits{G.}},
\oauthor{\bsnm{Cardoso}, \binits{F.}},
\oauthor{\bsnm{Cresci}, \binits{S.}},
\oauthor{\bsnm{Giordano}, \binits{S.}},
\oauthor{\bsnm{Luceri}, \binits{L.}}:
Tracking fringe and coordinated activity on twitter leading up to the us
  capitol attack.
arXiv preprint arXiv:2302.04450
(2023)
\end{botherref}
\endbibitem

%%% 8
\bibitem{ribeiro2018characterizing}
\begin{bchapter}
\bauthor{\bsnm{Ribeiro}, \binits{M.H.}},
\bauthor{\bsnm{Calais}, \binits{P.H.}},
\bauthor{\bsnm{Santos}, \binits{Y.A.}},
\bauthor{\bsnm{Almeida}, \binits{V.A.}},
\bauthor{\bsnm{Meira~Jr}, \binits{W.}}:
\bctitle{Characterizing and detecting hateful users on twitter}.
In: \bbtitle{Twelfth International AAAI Conference on Web and Social Media}
(\byear{2018})
\end{bchapter}
\endbibitem

%%% 9
\bibitem{bovet2019influence}
\begin{barticle}
\bauthor{\bsnm{Bovet}, \binits{A.}},
\bauthor{\bsnm{Makse}, \binits{H.A.}}:
\batitle{Influence of fake news in twitter during the 2016 us presidential
  election}.
\bjtitle{Nature communications}
\bvolume{10}(\bissue{1}),
\bfpage{1}--\blpage{14}
(\byear{2019})
\end{barticle}
\endbibitem

%%% 10
\bibitem{sharma2022characterizing}
\begin{bchapter}
\bauthor{\bsnm{Sharma}, \binits{K.}},
\bauthor{\bsnm{Ferrara}, \binits{E.}},
\bauthor{\bsnm{Liu}, \binits{Y.}}:
\bctitle{Characterizing online engagement with disinformation and conspiracies
  in the 2020 us presidential election}.
In: \bbtitle{16th International AAAI Conference on Web and Social Media}
(\byear{2022})
\end{bchapter}
\endbibitem

%%% 11
\bibitem{cinelli2020covid}
\begin{barticle}
\bauthor{\bsnm{Cinelli}, \binits{M.}},
\bauthor{\bsnm{Quattrociocchi}, \binits{W.}},
\bauthor{\bsnm{Galeazzi}, \binits{A.}},
\bauthor{\bsnm{Valensise}, \binits{C.M.}},
\bauthor{\bsnm{Brugnoli}, \binits{E.}},
\bauthor{\bsnm{Schmidt}, \binits{A.L.}},
\bauthor{\bsnm{Zola}, \binits{P.}},
\bauthor{\bsnm{Zollo}, \binits{F.}},
\bauthor{\bsnm{Scala}, \binits{A.}}:
\batitle{The covid-19 social media infodemic}.
\bjtitle{Scientific reports}
\bvolume{10}(\bissue{1}),
\bfpage{1}--\blpage{10}
(\byear{2020})
\end{barticle}
\endbibitem

%%% 12
\bibitem{gallotti2020assessing}
\begin{barticle}
\bauthor{\bsnm{Gallotti}, \binits{R.}},
\bauthor{\bsnm{Valle}, \binits{F.}},
\bauthor{\bsnm{Castaldo}, \binits{N.}},
\bauthor{\bsnm{Sacco}, \binits{P.}},
\bauthor{\bsnm{De~Domenico}, \binits{M.}}:
\batitle{Assessing the risks of ‘infodemics’ in response to covid-19
  epidemics}.
\bjtitle{Nature human behaviour}
\bvolume{4}(\bissue{12}),
\bfpage{1285}--\blpage{1293}
(\byear{2020})
\end{barticle}
\endbibitem

%%% 13
\bibitem{yang2021covid}
\begin{barticle}
\bauthor{\bsnm{Yang}, \binits{K.-C.}},
\bauthor{\bsnm{Pierri}, \binits{F.}},
\bauthor{\bsnm{Hui}, \binits{P.-M.}},
\bauthor{\bsnm{Axelrod}, \binits{D.}},
\bauthor{\bsnm{Torres-Lugo}, \binits{C.}},
\bauthor{\bsnm{Bryden}, \binits{J.}},
\bauthor{\bsnm{Menczer}, \binits{F.}}:
\batitle{The covid-19 infodemic: Twitter versus facebook}.
\bjtitle{Big Data \& Society}
\bvolume{8}(\bissue{1}),
\bfpage{20539517211013861}
(\byear{2021})
\end{barticle}
\endbibitem

%%% 14
\bibitem{chen2021covid}
\begin{botherref}
\oauthor{\bsnm{Chen}, \binits{E.}},
\oauthor{\bsnm{Chang}, \binits{H.}},
\oauthor{\bsnm{Rao}, \binits{A.}},
\oauthor{\bsnm{Lerman}, \binits{K.}},
\oauthor{\bsnm{Cowan}, \binits{G.}},
\oauthor{\bsnm{Ferrara}, \binits{E.}}:
Covid-19 misinformation and the 2020 us presidential election.
The Harvard Kennedy School Misinformation Review
\textbf{1}(7)
(2021)
\end{botherref}
\endbibitem

%%% 15
\bibitem{jiang2021social}
\begin{barticle}
\bauthor{\bsnm{Jiang}, \binits{J.}},
\bauthor{\bsnm{Ren}, \binits{X.}},
\bauthor{\bsnm{Ferrara}, \binits{E.}}, \betal:
\batitle{Social media polarization and echo chambers in the context of
  covid-19: Case study}.
\bjtitle{JMIRx med}
\bvolume{2}(\bissue{3}),
\bfpage{29570}
(\byear{2021})
\end{barticle}
\endbibitem

%%% 16
\bibitem{rao2021political}
\begin{barticle}
\bauthor{\bsnm{Rao}, \binits{A.}},
\bauthor{\bsnm{Morstatter}, \binits{F.}},
\bauthor{\bsnm{Hu}, \binits{M.}},
\bauthor{\bsnm{Chen}, \binits{E.}},
\bauthor{\bsnm{Burghardt}, \binits{K.}},
\bauthor{\bsnm{Ferrara}, \binits{E.}},
\bauthor{\bsnm{Lerman}, \binits{K.}}:
\batitle{Political partisanship and antiscience attitudes in online discussions
  about covid-19: Twitter content analysis}.
\bjtitle{Journal of medical Internet research}
\bvolume{23}(\bissue{6}),
\bfpage{26692}
(\byear{2021})
\end{barticle}
\endbibitem

%%% 17
\bibitem{chen2022charting}
\begin{barticle}
\bauthor{\bsnm{Chen}, \binits{E.}},
\bauthor{\bsnm{Jiang}, \binits{J.}},
\bauthor{\bsnm{Chang}, \binits{H.-C.H.}},
\bauthor{\bsnm{Muric}, \binits{G.}},
\bauthor{\bsnm{Ferrara}, \binits{E.}}:
\batitle{Charting the information and misinformation landscape to characterize
  misinfodemics on social media: Covid-19 infodemiology study at a planetary
  scale}.
\bjtitle{Jmir Infodemiology}
\bvolume{2}(\bissue{1}),
\bfpage{32378}
(\byear{2022})
\end{barticle}
\endbibitem

%%% 18
\bibitem{nogara2022disinformation}
\begin{bchapter}
\bauthor{\bsnm{Nogara}, \binits{G.}},
\bauthor{\bsnm{Vishnuprasad}, \binits{P.S.}},
\bauthor{\bsnm{Cardoso}, \binits{F.}},
\bauthor{\bsnm{Ayoub}, \binits{O.}},
\bauthor{\bsnm{Giordano}, \binits{S.}},
\bauthor{\bsnm{Luceri}, \binits{L.}}:
\bctitle{The disinformation dozen: An exploratory analysis of covid-19
  disinformation proliferation on twitter}.
In: \bbtitle{14th ACM Web Science Conference 2022},
pp. \bfpage{348}--\blpage{358}
(\byear{2022})
\end{bchapter}
\endbibitem

%%% 19
\bibitem{jhaver2021evaluating}
\begin{barticle}
\bauthor{\bsnm{Jhaver}, \binits{S.}},
\bauthor{\bsnm{Boylston}, \binits{C.}},
\bauthor{\bsnm{Yang}, \binits{D.}},
\bauthor{\bsnm{Bruckman}, \binits{A.}}:
\batitle{Evaluating the effectiveness of deplatforming as a moderation strategy
  on twitter}.
\bjtitle{Proceedings of the ACM on Human-Computer Interaction}
\bvolume{5}(\bissue{CSCW2}),
\bfpage{1}--\blpage{30}
(\byear{2021})
\end{barticle}
\endbibitem

%%% 20
\bibitem{chang2022comparative}
\begin{botherref}
\oauthor{\bsnm{Chang}, \binits{H.-C.H.}},
\oauthor{\bsnm{Ferrara}, \binits{E.}}:
Comparative analysis of social bots and humans during the covid-19 pandemic.
Journal of Computational Social Science,
1409--1425
(2022)
\end{botherref}
\endbibitem

%%% 21
\bibitem{wang2023identifying}
\begin{bchapter}
\bauthor{\bsnm{Wang}, \binits{E.}},
\bauthor{\bsnm{Luceri}, \binits{L.}},
\bauthor{\bsnm{Pierri}, \binits{F.}},
\bauthor{\bsnm{Ferrara}, \binits{E.}}:
\bctitle{Identifying and characterizing behavioral classes of radicalization
  within the qanon conspiracy on twitter}.
In: \bbtitle{17th International Conference on Web and Social Media}
(\byear{2023})
\end{bchapter}
\endbibitem

%%% 22
\bibitem{luceri2021social}
\begin{botherref}
\oauthor{\bsnm{Luceri}, \binits{L.}},
\oauthor{\bsnm{Cresci}, \binits{S.}},
\oauthor{\bsnm{Giordano}, \binits{S.}}:
Social media against society.
The Internet and the 2020 Campaign,
1
(2021)
\end{botherref}
\endbibitem

%%% 23
\bibitem{ali2021understanding}
\begin{bchapter}
\bauthor{\bsnm{Ali}, \binits{S.}},
\bauthor{\bsnm{Saeed}, \binits{M.H.}},
\bauthor{\bsnm{Aldreabi}, \binits{E.}},
\bauthor{\bsnm{Blackburn}, \binits{J.}},
\bauthor{\bsnm{De~Cristofaro}, \binits{E.}},
\bauthor{\bsnm{Zannettou}, \binits{S.}},
\bauthor{\bsnm{Stringhini}, \binits{G.}}:
\bctitle{Understanding the effect of deplatforming on social networks}.
In: \bbtitle{13th ACM Web Science Conference 2021},
pp. \bfpage{187}--\blpage{195}
(\byear{2021})
\end{bchapter}
\endbibitem

%%% 24
\bibitem{papakyriakopoulos2022impact}
\begin{bchapter}
\bauthor{\bsnm{Papakyriakopoulos}, \binits{O.}},
\bauthor{\bsnm{Goodman}, \binits{E.}}:
\bctitle{The impact of twitter labels on misinformation spread and user
  engagement: Lessons from trump’s election tweets}.
In: \bbtitle{Proceedings of the ACM Web Conference},
pp. \bfpage{2541}--\blpage{2551}
(\byear{2022})
\end{bchapter}
\endbibitem

%%% 25
\bibitem{zannettou2021won}
\begin{bchapter}
\bauthor{\bsnm{Zannettou}, \binits{S.}}:
\bctitle{" i won the election!": An empirical analysis of soft moderation
  interventions on twitter}.
In: \bbtitle{Proceedings of the International AAAI Conference on Web and Social
  Media},
vol. \bseriesno{15},
pp. \bfpage{865}--\blpage{876}
(\byear{2021})
\end{bchapter}
\endbibitem

%%% 26
\bibitem{le2019postmortem}
\begin{bchapter}
\bauthor{\bsnm{Le}, \binits{H.}},
\bauthor{\bsnm{Boynton}, \binits{G.}},
\bauthor{\bsnm{Shafiq}, \binits{Z.}},
\bauthor{\bsnm{Srinivasan}, \binits{P.}}:
\bctitle{A postmortem of suspended twitter accounts in the 2016 us presidential
  election}.
In: \bbtitle{2019 IEEE/ACM International ASONAM Conference},
pp. \bfpage{258}--\blpage{265}
(\byear{2019})
\end{bchapter}
\endbibitem

%%% 27
\bibitem{toraman2022blacklivesmatter}
\begin{bchapter}
\bauthor{\bsnm{Toraman}, \binits{C.}},
\bauthor{\bsnm{{\c{S}}ahinu{\c{c}}}, \binits{F.}},
\bauthor{\bsnm{Yilmaz}, \binits{E.H.}}:
\bctitle{Blacklivesmatter 2020: An analysis of deleted and suspended users in
  twitter}.
In: \bbtitle{14th ACM Web Science Conference 2022},
pp. \bfpage{290}--\blpage{295}
(\byear{2022})
\end{bchapter}
\endbibitem

%%% 28
\bibitem{majo2021role}
\begin{botherref}
\oauthor{\bsnm{Maj{\'o}-V{\'a}zquez}, \binits{S.}},
\oauthor{\bsnm{Congosto}, \binits{M.}},
\oauthor{\bsnm{Nicholls}, \binits{T.}},
\oauthor{\bsnm{Nielsen}, \binits{R.K.}}:
The role of suspended accounts in political discussion on social media:
  Analysis of the 2017 french, uk and german elections.
Social Media+ Society
(2021)
\end{botherref}
\endbibitem

%%% 29
\bibitem{guo2022large}
\begin{bchapter}
\bauthor{\bsnm{Guo}, \binits{X.}},
\bauthor{\bsnm{Vosoughi}, \binits{S.}}:
\bctitle{A large-scale longitudinal multimodal dataset of state-backed
  information operations on twitter}.
In: \bbtitle{Proceedings of the International AAAI Conference on Web and Social
  Media}
(\byear{2022})
\end{bchapter}
\endbibitem

%%% 30
\bibitem{badawy2018analyzing}
\begin{bchapter}
\bauthor{\bsnm{Badawy}, \binits{A.}},
\bauthor{\bsnm{Ferrara}, \binits{E.}},
\bauthor{\bsnm{Lerman}, \binits{K.}}:
\bctitle{Analyzing the digital traces of political manipulation: The 2016
  russian interference twitter campaign}.
In: \bbtitle{2018 IEEE/ACM International ASONAM Conference},
pp. \bfpage{258}--\blpage{265}
(\byear{2018})
\end{bchapter}
\endbibitem

%%% 31
\bibitem{stella2018bots}
\begin{barticle}
\bauthor{\bsnm{Stella}, \binits{M.}},
\bauthor{\bsnm{Ferrara}, \binits{E.}},
\bauthor{\bsnm{De~Domenico}, \binits{M.}}:
\batitle{Bots increase exposure to negative and inflammatory content in online
  social systems}.
\bjtitle{Proceedings of the National Academy of Sciences}
\bvolume{115}(\bissue{49}),
\bfpage{12435}--\blpage{12440}
(\byear{2018})
\end{barticle}
\endbibitem

%%% 32
\bibitem{jiang2020political}
\begin{barticle}
\bauthor{\bsnm{Jiang}, \binits{J.}},
\bauthor{\bsnm{Chen}, \binits{E.}},
\bauthor{\bsnm{Yan}, \binits{S.}},
\bauthor{\bsnm{Lerman}, \binits{K.}},
\bauthor{\bsnm{Ferrara}, \binits{E.}}:
\batitle{Political polarization drives online conversations about covid-19 in
  the united states}.
\bjtitle{Human Behavior and Emerging Technologies}
\bvolume{2}(\bissue{3}),
\bfpage{200}--\blpage{211}
(\byear{2020})
\end{barticle}
\endbibitem

%%% 33
\bibitem{osomewp1}
\begin{botherref}
\oauthor{\bsnm{{IU Observatory on Social Media}}}:
Suspicious Twitter Activity Around the Russian Invasion of Ukraine
\end{botherref}
\endbibitem

%%% 34
\bibitem{osomewp2}
\begin{botherref}
\oauthor{\bsnm{{IU Observatory on Social Media}}}:
Analysis of twitter accounts created around the invasion of ukraine
(2022)
\end{botherref}
\endbibitem

%%% 35
\bibitem{ferrara2017disinformation}
\begin{botherref}
\oauthor{\bsnm{Ferrara}, \binits{E.}}:
Disinformation and social bot operations in the run up to the 2017 french
  presidential election.
First Monday
\textbf{22}(8)
(2017)
\end{botherref}
\endbibitem

%%% 36
\bibitem{stringhini2010detecting}
\begin{bchapter}
\bauthor{\bsnm{Stringhini}, \binits{G.}},
\bauthor{\bsnm{Kruegel}, \binits{C.}},
\bauthor{\bsnm{Vigna}, \binits{G.}}:
\bctitle{Detecting spammers on social networks}.
In: \bbtitle{Proceedings of the 26th Annual Computer Security Applications
  Conference},
pp. \bfpage{1}--\blpage{9}
(\byear{2010})
\end{bchapter}
\endbibitem

%%% 37
\bibitem{yang2012analyzing}
\begin{bchapter}
\bauthor{\bsnm{Yang}, \binits{C.}},
\bauthor{\bsnm{Harkreader}, \binits{R.}},
\bauthor{\bsnm{Zhang}, \binits{J.}},
\bauthor{\bsnm{Shin}, \binits{S.}},
\bauthor{\bsnm{Gu}, \binits{G.}}:
\bctitle{Analyzing spammers' social networks for fun and profit: a case study
  of cyber criminal ecosystem on twitter}.
In: \bbtitle{Proceedings of the 21st International Conference on World Wide
  Web},
pp. \bfpage{71}--\blpage{80}
(\byear{2012})
\end{bchapter}
\endbibitem

%%% 38
\bibitem{ferrara2019history}
\begin{barticle}
\bauthor{\bsnm{Ferrara}, \binits{E.}}:
\batitle{The history of digital spam}.
\bjtitle{Communications of the ACM}
\bvolume{62}(\bissue{8}),
\bfpage{82}--\blpage{91}
(\byear{2019})
\end{barticle}
\endbibitem

%%% 39
\bibitem{ferrara2022twitter}
\begin{botherref}
\oauthor{\bsnm{Ferrara}, \binits{E.}}:
Twitter spam and false accounts prevalence, detection, and characterization: A
  survey.
First Monday
\textbf{27}(12)
(2022)
\end{botherref}
\endbibitem

%%% 40
\bibitem{luceri2020detecting}
\begin{bchapter}
\bauthor{\bsnm{Luceri}, \binits{L.}},
\bauthor{\bsnm{Giordano}, \binits{S.}},
\bauthor{\bsnm{Ferrara}, \binits{E.}}:
\bctitle{Detecting troll behavior via inverse reinforcement learning: A case
  study of russian trolls in the 2016 us election}.
In: \bbtitle{Proceedings of the International AAAI Conference on Web and Social
  Media},
vol. \bseriesno{14},
pp. \bfpage{417}--\blpage{427}
(\byear{2020})
\end{bchapter}
\endbibitem

%%% 41
\bibitem{mazza2022investigating}
\begin{barticle}
\bauthor{\bsnm{Mazza}, \binits{M.}},
\bauthor{\bsnm{Avvenuti}, \binits{M.}},
\bauthor{\bsnm{Cresci}, \binits{S.}},
\bauthor{\bsnm{Tesconi}, \binits{M.}}:
\batitle{Investigating the difference between trolls, social bots, and humans
  on twitter}.
\bjtitle{Computer Communications}
\bvolume{196},
\bfpage{23}--\blpage{36}
(\byear{2022})
\end{barticle}
\endbibitem

%%% 42
\bibitem{addawood2019linguistic}
\begin{bchapter}
\bauthor{\bsnm{Addawood}, \binits{A.}},
\bauthor{\bsnm{Badawy}, \binits{A.}},
\bauthor{\bsnm{Lerman}, \binits{K.}},
\bauthor{\bsnm{Ferrara}, \binits{E.}}:
\bctitle{Linguistic cues to deception: Identifying political trolls on social
  media}.
In: \bbtitle{Proceedings of the International AAAI Conference on Web and Social
  Media},
vol. \bseriesno{13},
pp. \bfpage{15}--\blpage{25}
(\byear{2019})
\end{bchapter}
\endbibitem

%%% 43
\bibitem{chowdhury2020twitter}
\begin{bchapter}
\bauthor{\bsnm{Chowdhury}, \binits{F.A.}},
\bauthor{\bsnm{Allen}, \binits{L.}},
\bauthor{\bsnm{Yousuf}, \binits{M.}},
\bauthor{\bsnm{Mueen}, \binits{A.}}:
\bctitle{On twitter purge: a retrospective analysis of suspended users}.
In: \bbtitle{Companion Proceedings of the Web Conference},
pp. \bfpage{371}--\blpage{378}
(\byear{2020})
\end{bchapter}
\endbibitem

%%% 44
\bibitem{chowdhury2021examining}
\begin{bchapter}
\bauthor{\bsnm{Chowdhury}, \binits{F.A.}},
\bauthor{\bsnm{Saha}, \binits{D.}},
\bauthor{\bsnm{Hasan}, \binits{M.R.}},
\bauthor{\bsnm{Saha}, \binits{K.}},
\bauthor{\bsnm{Mueen}, \binits{A.}}:
\bctitle{Examining factors associated with twitter account suspension following
  the 2020 us presidential election}.
In: \bbtitle{Proceedings of the 2021 IEEE/ACM International Conference on
  Advances in Social Networks Analysis and Mining},
pp. \bfpage{607}--\blpage{612}
(\byear{2021})
\end{bchapter}
\endbibitem

%%% 45
\bibitem{seyler2021textual}
\begin{bchapter}
\bauthor{\bsnm{Seyler}, \binits{D.}},
\bauthor{\bsnm{Tan}, \binits{S.}},
\bauthor{\bsnm{Li}, \binits{D.}},
\bauthor{\bsnm{Zhang}, \binits{J.}},
\bauthor{\bsnm{Li}, \binits{P.}}:
\bctitle{Textual analysis and timely detection of suspended social media
  accounts.}
In: \bbtitle{ICWSM},
pp. \bfpage{644}--\blpage{655}
(\byear{2021})
\end{bchapter}
\endbibitem

%%% 46
\bibitem{caprolu2022characterizing}
\begin{botherref}
\oauthor{\bsnm{Caprolu}, \binits{M.}},
\oauthor{\bsnm{Sadighian}, \binits{A.}},
\oauthor{\bsnm{Di~Pietro}, \binits{R.}}:
Characterizing the 2022 russo-ukrainian conflict through the lenses of
  aspect-based sentiment analysis: Dataset, methodology, and preliminary
  findings.
arXiv:2208.04903
(2022)
\end{botherref}
\endbibitem

%%% 47
\bibitem{park2022voynaslov}
\begin{botherref}
\oauthor{\bsnm{Park}, \binits{C.Y.}},
\oauthor{\bsnm{Mendelsohn}, \binits{J.}},
\oauthor{\bsnm{Field}, \binits{A.}},
\oauthor{\bsnm{Tsvetkov}, \binits{Y.}}:
Voynaslov: A data set of russian social media activity during the 2022
  ukraine-russia war.
arXiv:2205.12382
(2022)
\end{botherref}
\endbibitem

%%% 48
\bibitem{hanley2022happenstance}
\begin{botherref}
\oauthor{\bsnm{Hanley}, \binits{H.W.}},
\oauthor{\bsnm{Kumar}, \binits{D.}},
\oauthor{\bsnm{Durumeric}, \binits{Z.}}:
Happenstance: Utilizing semantic search to track russian state media narratives
  about the russo-ukrainian war on reddit.
arXiv:2205.14484
(2022)
\end{botherref}
\endbibitem

%%% 49
\bibitem{hanley2022special}
\begin{botherref}
\oauthor{\bsnm{Hanley}, \binits{H.W.}},
\oauthor{\bsnm{Kumar}, \binits{D.}},
\oauthor{\bsnm{Durumeric}, \binits{Z.}}:
" a special operation": A quantitative approach to dissecting and comparing
  different media ecosystems' coverage of the russo-ukrainian war.
arXiv:2210.03016
(2022)
\end{botherref}
\endbibitem

%%% 50
\bibitem{geissler2022russian}
\begin{botherref}
\oauthor{\bsnm{Geissler}, \binits{D.}},
\oauthor{\bsnm{B{\"a}r}, \binits{D.}},
\oauthor{\bsnm{Pr{\"o}llochs}, \binits{N.}},
\oauthor{\bsnm{Feuerriegel}, \binits{S.}}:
Russian propaganda on social media during the 2022 invasion of ukraine.
arXiv:2211.04154
(2022)
\end{botherref}
\endbibitem

%%% 51
\bibitem{pierri2023propaganda}
\begin{bchapter}
\bauthor{\bsnm{Pierri}, \binits{F.}},
\bauthor{\bsnm{Luceri}, \binits{L.}},
\bauthor{\bsnm{Jindal}, \binits{N.}},
\bauthor{\bsnm{Ferrara}, \binits{E.}}:
\bctitle{Propaganda and misinformation on facebook and twitter during the
  russian invasion of ukraine}.
In: \bbtitle{WebSci’23 -- 15th ACM Web Science Conference}
(\byear{2023})
\end{bchapter}
\endbibitem

%%% 52
\bibitem{cinelli2020limited}
\begin{barticle}
\bauthor{\bsnm{Cinelli}, \binits{M.}},
\bauthor{\bsnm{Cresci}, \binits{S.}},
\bauthor{\bsnm{Galeazzi}, \binits{A.}},
\bauthor{\bsnm{Quattrociocchi}, \binits{W.}},
\bauthor{\bsnm{Tesconi}, \binits{M.}}:
\batitle{The limited reach of fake news on twitter during 2019 european
  elections}.
\bjtitle{PloS one}
\bvolume{15}(\bissue{6}),
\bfpage{0234689}
(\byear{2020})
\end{barticle}
\endbibitem

%%% 53
\bibitem{grinberg2019fake}
\begin{barticle}
\bauthor{\bsnm{Grinberg}, \binits{N.}},
\bauthor{\bsnm{Joseph}, \binits{K.}},
\bauthor{\bsnm{Friedland}, \binits{L.}},
\bauthor{\bsnm{Swire-Thompson}, \binits{B.}},
\bauthor{\bsnm{Lazer}, \binits{D.}}:
\batitle{Fake news on twitter during the 2016 us presidential election}.
\bjtitle{Science}
\bvolume{363}(\bissue{6425}),
\bfpage{374}--\blpage{378}
(\byear{2019})
\end{barticle}
\endbibitem

%%% 54
\bibitem{ferrara2020bots}
\begin{botherref}
\oauthor{\bsnm{Ferrara}, \binits{E.}}:
Bots, elections, and social media: a brief overview.
Disinformation, Misinformation, and Fake News in Social Media,
95--114
(2020)
\end{botherref}
\endbibitem

%%% 55
\bibitem{jiang2023retweet}
\begin{bchapter}
\bauthor{\bsnm{Jiang}, \binits{J.}},
\bauthor{\bsnm{Ren}, \binits{X.}},
\bauthor{\bsnm{Ferrara}, \binits{E.}}:
\bctitle{Retweet-bert: Political leaning detection using language features and
  information diffusion on social networks}.
In: \bbtitle{17th International AAAI Conference on Web and Social Media}
(\byear{2023})
\end{bchapter}
\endbibitem

%%% 56
\bibitem{abdine2022political}
\begin{botherref}
\oauthor{\bsnm{Abdine}, \binits{H.}},
\oauthor{\bsnm{Guo}, \binits{Y.}},
\oauthor{\bsnm{Rennard}, \binits{V.}},
\oauthor{\bsnm{Vazirgiannis}, \binits{M.}}:
Political communities on twitter: Case study of the 2022 french presidential
  election.
arXiv:2204.07436
(2022)
\end{botherref}
\endbibitem

%%% 57
\bibitem{chen2022tweets}
\begin{bchapter}
\bauthor{\bsnm{Chen}, \binits{E.}},
\bauthor{\bsnm{Ferrara}, \binits{E.}}:
\bctitle{Tweets in time of conflict: A public dataset tracking the twitter
  discourse on the war between ukraine and russia}.
In: \bbtitle{17th International AAAI Conference on Web and Social Media
  (ICWSM'23)}
(\byear{2023})
\end{bchapter}
\endbibitem

%%% 58
\bibitem{munch}
\begin{botherref}
\oauthor{\bsnm{Munch}, \binits{F.V.}},
\oauthor{\bsnm{Kessling}, \binits{P.}}:
Ukraine Twitter Data.
doi:\doiurl{10.17605/OSF.IO/RTQXN}.
\url{osf.io/rtqxn}
\end{botherref}
\endbibitem

%%% 59
\bibitem{deverna2021covaxxy}
\begin{botherref}
\oauthor{\bsnm{DeVerna}, \binits{M.}},
\oauthor{\bsnm{Pierri}, \binits{F.}},
\oauthor{\bsnm{Truong}, \binits{B.}},
\oauthor{\bsnm{Bollenbacher}, \binits{J.}},
\oauthor{\bsnm{Axelrod}, \binits{D.}},
\oauthor{\bsnm{Loynes}, \binits{N.}},
\oauthor{\bsnm{Torres-Lugo}, \binits{C.}},
\oauthor{\bsnm{Yang}, \binits{K.-C.}},
\oauthor{\bsnm{Menczer}, \binits{F.}},
\oauthor{\bsnm{Bryden}, \binits{J.}}:
Covaxxy: A global collection of english twitter posts about covid-19 vaccines.
Proceedings of the International AAAI Conference on Web and Social Media
(2021)
\end{botherref}
\endbibitem

%%% 60
\bibitem{morstatter2013sample}
\begin{bchapter}
\bauthor{\bsnm{Morstatter}, \binits{F.}},
\bauthor{\bsnm{Pfeffer}, \binits{J.}},
\bauthor{\bsnm{Liu}, \binits{H.}},
\bauthor{\bsnm{Carley}, \binits{K.}}:
\bctitle{Is the sample good enough? comparing data from twitter's streaming api
  with twitter's firehose}.
In: \bbtitle{Proceedings of the International AAAI Conference on Web and Social
  Media},
vol. \bseriesno{7},
pp. \bfpage{400}--\blpage{408}
(\byear{2013})
\end{bchapter}
\endbibitem

%%% 61
\bibitem{arora2023detecting}
\begin{botherref}
\oauthor{\bsnm{Arora}, \binits{A.}},
\oauthor{\bsnm{Nakov}, \binits{P.}},
\oauthor{\bsnm{Hardalov}, \binits{M.}},
\oauthor{\bsnm{Sarwar}, \binits{S.M.}},
\oauthor{\bsnm{Nayak}, \binits{V.}},
\oauthor{\bsnm{Dinkov}, \binits{Y.}},
\oauthor{\bsnm{Zlatkova}, \binits{D.}},
\oauthor{\bsnm{Dent}, \binits{K.}},
\oauthor{\bsnm{Bhatawdekar}, \binits{A.}},
\oauthor{\bsnm{Bouchard}, \binits{G.}}, et al.:
Detecting harmful content on online platforms: What platforms need vs. where
  research efforts go.
ACM Computing Surveys
(2023)
\end{botherref}
\endbibitem

%%% 62
\bibitem{torres2022manipulating}
\begin{bchapter}
\bauthor{\bsnm{Torres-Lugo}, \binits{C.}},
\bauthor{\bsnm{Pote}, \binits{M.}},
\bauthor{\bsnm{Nwala}, \binits{A.C.}},
\bauthor{\bsnm{Menczer}, \binits{F.}}:
\bctitle{Manipulating twitter through deletions}.
In: \bbtitle{Proceedings of the International AAAI Conference on Web and Social
  Media},
vol. \bseriesno{16},
pp. \bfpage{1029}--\blpage{1039}
(\byear{2022})
\end{bchapter}
\endbibitem

%%% 63
\bibitem{luceri2019red}
\begin{bchapter}
\bauthor{\bsnm{Luceri}, \binits{L.}},
\bauthor{\bsnm{Deb}, \binits{A.}},
\bauthor{\bsnm{Badawy}, \binits{A.}},
\bauthor{\bsnm{Ferrara}, \binits{E.}}:
\bctitle{Red bots do it better: Comparative analysis of social bot partisan
  behavior}.
In: \bbtitle{Companion Proceedings of the 2019 World Wide Web Conference},
pp. \bfpage{1007}--\blpage{1012}
(\byear{2019})
\end{bchapter}
\endbibitem

%%% 64
\bibitem{ye2023online}
\begin{bchapter}
\bauthor{\bsnm{Ye}, \binits{J.}},
\bauthor{\bsnm{Jindal}, \binits{N.}},
\bauthor{\bsnm{Pierri}, \binits{F.}},
\bauthor{\bsnm{Luceri}, \binits{L.}}:
\bctitle{Online networks of support in distressed environments: Solidarity and
  mobilization during the russian invasion of ukraine}.
In: \bbtitle{Companion Proceedings of ICWSM 2023}
(\byear{2023})
\end{bchapter}
\endbibitem

%%% 65
\bibitem{nizzoli2020charting}
\begin{barticle}
\bauthor{\bsnm{Nizzoli}, \binits{L.}},
\bauthor{\bsnm{Tardelli}, \binits{S.}},
\bauthor{\bsnm{Avvenuti}, \binits{M.}},
\bauthor{\bsnm{Cresci}, \binits{S.}},
\bauthor{\bsnm{Tesconi}, \binits{M.}},
\bauthor{\bsnm{Ferrara}, \binits{E.}}:
\batitle{Charting the landscape of online cryptocurrency manipulation}.
\bjtitle{IEEE Access}
\bvolume{8},
\bfpage{113230}--\blpage{113245}
(\byear{2020})
\end{barticle}
\endbibitem

%%% 66
\bibitem{nghiem2021detecting}
\begin{barticle}
\bauthor{\bsnm{Nghiem}, \binits{H.}},
\bauthor{\bsnm{Muric}, \binits{G.}},
\bauthor{\bsnm{Morstatter}, \binits{F.}},
\bauthor{\bsnm{Ferrara}, \binits{E.}}:
\batitle{Detecting cryptocurrency pump-and-dump frauds using market and social
  signals}.
\bjtitle{Expert Systems with Applications}
\bvolume{182},
\bfpage{115284}
(\byear{2021})
\end{barticle}
\endbibitem

%%% 67
\bibitem{founta2018large}
\begin{bchapter}
\bauthor{\bsnm{Founta}, \binits{A.M.}},
\bauthor{\bsnm{Djouvas}, \binits{C.}},
\bauthor{\bsnm{Chatzakou}, \binits{D.}},
\bauthor{\bsnm{Leontiadis}, \binits{I.}},
\bauthor{\bsnm{Blackburn}, \binits{J.}},
\bauthor{\bsnm{Stringhini}, \binits{G.}},
\bauthor{\bsnm{Vakali}, \binits{A.}},
\bauthor{\bsnm{Sirivianos}, \binits{M.}},
\bauthor{\bsnm{Kourtellis}, \binits{N.}}:
\bctitle{Large scale crowdsourcing and characterization of twitter abusive
  behavior}.
In: \bbtitle{Twelfth International AAAI Conference on Web and Social Media}
(\byear{2018})
\end{bchapter}
\endbibitem

%%% 68
\bibitem{yang2022botometer}
\begin{barticle}
\bauthor{\bsnm{Yang}, \binits{K.-C.}},
\bauthor{\bsnm{Ferrara}, \binits{E.}},
\bauthor{\bsnm{Menczer}, \binits{F.}}:
\batitle{Botometer 101: Social bot practicum for computational social
  scientists}.
\bjtitle{Journal of computational social science}
\bvolume{5},
\bfpage{1511}--\blpage{1528}
(\byear{2022})
\end{barticle}
\endbibitem

%%% 69
\bibitem{ferrara2020types}
\begin{botherref}
\oauthor{\bsnm{Ferrara}, \binits{E.}}:
What types of covid-19 conspiracies are populated by twitter bots?
arXiv preprint arXiv:2004.09531
(2020)
\end{botherref}
\endbibitem

%%% 70
\bibitem{luceri2021down}
\begin{botherref}
\oauthor{\bsnm{Luceri}, \binits{L.}},
\oauthor{\bsnm{Cardoso}, \binits{F.}},
\oauthor{\bsnm{Giordano}, \binits{S.}}:
Down the bot hole: Actionable insights from a one-year analysis of bot activity
  on twitter.
First Monday
(2021)
\end{botherref}
\endbibitem

\end{thebibliography}

\newcommand{\BMCxmlcomment}[1]{}

\BMCxmlcomment{

<refgrp>

<bibl id="B1">
  <title><p>Facebook use and political participation: The impact of exposure to
  shared political information, connections with public political actors, and
  network structural heterogeneity</p></title>
  <aug>
    <au><snm>Tang</snm><fnm>G</fnm></au>
    <au><snm>Lee</snm><fnm>FL</fnm></au>
  </aug>
  <source>Social science computer review</source>
  <publisher>Sage Publications Sage CA: Los Angeles, CA</publisher>
  <pubdate>2013</pubdate>
  <volume>31</volume>
  <issue>6</issue>
  <fpage>763</fpage>
  <lpage>-773</lpage>
</bibl>

<bibl id="B2">
  <title><p>Social network sites and political engagement: Exploring the impact
  of Facebook connections and uses on political protest and
  participation</p></title>
  <aug>
    <au><snm>Chan</snm><fnm>M</fnm></au>
  </aug>
  <source>Mass communication and society</source>
  <publisher>Taylor \& Francis</publisher>
  <pubdate>2016</pubdate>
  <volume>19</volume>
  <issue>4</issue>
  <fpage>430</fpage>
  <lpage>-451</lpage>
</bibl>

<bibl id="B3">
  <title><p>The rise of social bots</p></title>
  <aug>
    <au><snm>Ferrara</snm><fnm>E</fnm></au>
    <au><snm>Varol</snm><fnm>O</fnm></au>
    <au><snm>Davis</snm><fnm>C</fnm></au>
    <au><snm>Menczer</snm><fnm>F</fnm></au>
    <au><snm>Flammini</snm><fnm>A</fnm></au>
  </aug>
  <source>Communications of the ACM</source>
  <publisher>ACM New York, NY, USA</publisher>
  <pubdate>2016</pubdate>
  <volume>59</volume>
  <issue>7</issue>
  <fpage>96</fpage>
  <lpage>-104</lpage>
</bibl>

<bibl id="B4">
  <title><p>The science of fake news</p></title>
  <aug>
    <au><snm>Lazer</snm><fnm>DM</fnm></au>
    <au><snm>Baum</snm><fnm>MA</fnm></au>
    <au><snm>Benkler</snm><fnm>Y</fnm></au>
    <au><snm>Berinsky</snm><fnm>AJ</fnm></au>
    <au><snm>Greenhill</snm><fnm>KM</fnm></au>
    <au><snm>Menczer</snm><fnm>F</fnm></au>
    <au><snm>Metzger</snm><fnm>MJ</fnm></au>
    <au><snm>Nyhan</snm><fnm>B</fnm></au>
    <au><snm>Pennycook</snm><fnm>G</fnm></au>
    <au><snm>Rothschild</snm><fnm>D</fnm></au>
    <au><cnm>others</cnm></au>
  </aug>
  <source>Science</source>
  <publisher>American Association for the Advancement of Science</publisher>
  <pubdate>2018</pubdate>
  <volume>359</volume>
  <issue>6380</issue>
  <fpage>1094</fpage>
  <lpage>-1096</lpage>
</bibl>

<bibl id="B5">
  <title><p>Protecting elections from social media manipulation</p></title>
  <aug>
    <au><snm>Aral</snm><fnm>S</fnm></au>
    <au><snm>Eckles</snm><fnm>D</fnm></au>
  </aug>
  <source>Science</source>
  <publisher>American Association for the Advancement of Science</publisher>
  <pubdate>2019</pubdate>
  <volume>365</volume>
  <issue>6456</issue>
  <fpage>858</fpage>
  <lpage>-861</lpage>
</bibl>

<bibl id="B6">
  <title><p>Identifying Coordinated Accounts on Social Media through Hidden
  Influence and Group Behaviours</p></title>
  <aug>
    <au><snm>Sharma</snm><fnm>K</fnm></au>
    <au><snm>Zhang</snm><fnm>Y</fnm></au>
    <au><snm>Ferrara</snm><fnm>E</fnm></au>
    <au><snm>Liu</snm><fnm>Y</fnm></au>
  </aug>
  <source>KDD’21</source>
  <pubdate>2021</pubdate>
</bibl>

<bibl id="B7">
  <title><p>Tracking Fringe and Coordinated Activity on Twitter Leading Up To
  the US Capitol Attack</p></title>
  <aug>
    <au><snm>Suresh</snm><fnm>VP</fnm></au>
    <au><snm>Nogara</snm><fnm>G</fnm></au>
    <au><snm>Cardoso</snm><fnm>F</fnm></au>
    <au><snm>Cresci</snm><fnm>S</fnm></au>
    <au><snm>Giordano</snm><fnm>S</fnm></au>
    <au><snm>Luceri</snm><fnm>L</fnm></au>
  </aug>
  <source>arXiv preprint arXiv:2302.04450</source>
  <pubdate>2023</pubdate>
</bibl>

<bibl id="B8">
  <title><p>Characterizing and detecting hateful users on twitter</p></title>
  <aug>
    <au><snm>Ribeiro</snm><fnm>MH</fnm></au>
    <au><snm>Calais</snm><fnm>PH</fnm></au>
    <au><snm>Santos</snm><fnm>YA</fnm></au>
    <au><snm>Almeida</snm><fnm>VA</fnm></au>
    <au><snm>Meira Jr</snm><fnm>W</fnm></au>
  </aug>
  <source>Twelfth international AAAI conference on web and social
  media</source>
  <pubdate>2018</pubdate>
</bibl>

<bibl id="B9">
  <title><p>Influence of fake news in Twitter during the 2016 US presidential
  election</p></title>
  <aug>
    <au><snm>Bovet</snm><fnm>A</fnm></au>
    <au><snm>Makse</snm><fnm>HA</fnm></au>
  </aug>
  <source>Nature communications</source>
  <publisher>Nature Publishing Group</publisher>
  <pubdate>2019</pubdate>
  <volume>10</volume>
  <issue>1</issue>
  <fpage>1</fpage>
  <lpage>-14</lpage>
</bibl>

<bibl id="B10">
  <title><p>Characterizing Online Engagement with Disinformation and
  Conspiracies in the 2020 US Presidential Election</p></title>
  <aug>
    <au><snm>Sharma</snm><fnm>K</fnm></au>
    <au><snm>Ferrara</snm><fnm>E</fnm></au>
    <au><snm>Liu</snm><fnm>Y</fnm></au>
  </aug>
  <source>16th International AAAI Conference on Web and Social Media</source>
  <pubdate>2022</pubdate>
</bibl>

<bibl id="B11">
  <title><p>The COVID-19 social media infodemic</p></title>
  <aug>
    <au><snm>Cinelli</snm><fnm>M</fnm></au>
    <au><snm>Quattrociocchi</snm><fnm>W</fnm></au>
    <au><snm>Galeazzi</snm><fnm>A</fnm></au>
    <au><snm>Valensise</snm><fnm>CM</fnm></au>
    <au><snm>Brugnoli</snm><fnm>E</fnm></au>
    <au><snm>Schmidt</snm><fnm>AL</fnm></au>
    <au><snm>Zola</snm><fnm>P</fnm></au>
    <au><snm>Zollo</snm><fnm>F</fnm></au>
    <au><snm>Scala</snm><fnm>A</fnm></au>
  </aug>
  <source>Scientific reports</source>
  <publisher>Nature Publishing Group</publisher>
  <pubdate>2020</pubdate>
  <volume>10</volume>
  <issue>1</issue>
  <fpage>1</fpage>
  <lpage>-10</lpage>
</bibl>

<bibl id="B12">
  <title><p>Assessing the risks of ‘infodemics’ in response to COVID-19
  epidemics</p></title>
  <aug>
    <au><snm>Gallotti</snm><fnm>R</fnm></au>
    <au><snm>Valle</snm><fnm>F</fnm></au>
    <au><snm>Castaldo</snm><fnm>N</fnm></au>
    <au><snm>Sacco</snm><fnm>P</fnm></au>
    <au><snm>De Domenico</snm><fnm>M</fnm></au>
  </aug>
  <source>Nature human behaviour</source>
  <publisher>Nature Publishing Group</publisher>
  <pubdate>2020</pubdate>
  <volume>4</volume>
  <issue>12</issue>
  <fpage>1285</fpage>
  <lpage>-1293</lpage>
</bibl>

<bibl id="B13">
  <title><p>The covid-19 infodemic: Twitter versus facebook</p></title>
  <aug>
    <au><snm>Yang</snm><fnm>KC</fnm></au>
    <au><snm>Pierri</snm><fnm>F</fnm></au>
    <au><snm>Hui</snm><fnm>PM</fnm></au>
    <au><snm>Axelrod</snm><fnm>D</fnm></au>
    <au><snm>Torres Lugo</snm><fnm>C</fnm></au>
    <au><snm>Bryden</snm><fnm>J</fnm></au>
    <au><snm>Menczer</snm><fnm>F</fnm></au>
  </aug>
  <source>Big Data \& Society</source>
  <publisher>SAGE Publications Sage UK: London, England</publisher>
  <pubdate>2021</pubdate>
  <volume>8</volume>
  <issue>1</issue>
  <fpage>20539517211013861</fpage>
</bibl>

<bibl id="B14">
  <title><p>COVID-19 misinformation and the 2020 US presidential
  election</p></title>
  <aug>
    <au><snm>Chen</snm><fnm>E</fnm></au>
    <au><snm>Chang</snm><fnm>H</fnm></au>
    <au><snm>Rao</snm><fnm>A</fnm></au>
    <au><snm>Lerman</snm><fnm>K</fnm></au>
    <au><snm>Cowan</snm><fnm>G</fnm></au>
    <au><snm>Ferrara</snm><fnm>E</fnm></au>
  </aug>
  <source>The Harvard Kennedy School Misinformation Review</source>
  <publisher>Shorenstein Center for Media, Politics and Public Policy, at
  Harvard~…</publisher>
  <pubdate>2021</pubdate>
  <volume>1</volume>
  <issue>7</issue>
</bibl>

<bibl id="B15">
  <title><p>Social media polarization and echo chambers in the context of
  COVID-19: Case study</p></title>
  <aug>
    <au><snm>Jiang</snm><fnm>J</fnm></au>
    <au><snm>Ren</snm><fnm>X</fnm></au>
    <au><snm>Ferrara</snm><fnm>E</fnm></au>
    <au><cnm>others</cnm></au>
  </aug>
  <source>JMIRx med</source>
  <publisher>JMIR Publications Inc., Toronto, Canada</publisher>
  <pubdate>2021</pubdate>
  <volume>2</volume>
  <issue>3</issue>
  <fpage>e29570</fpage>
</bibl>

<bibl id="B16">
  <title><p>Political partisanship and antiscience attitudes in online
  discussions about COVID-19: Twitter content analysis</p></title>
  <aug>
    <au><snm>Rao</snm><fnm>A</fnm></au>
    <au><snm>Morstatter</snm><fnm>F</fnm></au>
    <au><snm>Hu</snm><fnm>M</fnm></au>
    <au><snm>Chen</snm><fnm>E</fnm></au>
    <au><snm>Burghardt</snm><fnm>K</fnm></au>
    <au><snm>Ferrara</snm><fnm>E</fnm></au>
    <au><snm>Lerman</snm><fnm>K</fnm></au>
  </aug>
  <source>Journal of medical Internet research</source>
  <publisher>JMIR Publications Toronto, Canada</publisher>
  <pubdate>2021</pubdate>
  <volume>23</volume>
  <issue>6</issue>
  <fpage>e26692</fpage>
</bibl>

<bibl id="B17">
  <title><p>Charting the information and misinformation landscape to
  characterize misinfodemics on social media: COVID-19 infodemiology study at a
  planetary scale</p></title>
  <aug>
    <au><snm>Chen</snm><fnm>E</fnm></au>
    <au><snm>Jiang</snm><fnm>J</fnm></au>
    <au><snm>Chang</snm><fnm>HCH</fnm></au>
    <au><snm>Muric</snm><fnm>G</fnm></au>
    <au><snm>Ferrara</snm><fnm>E</fnm></au>
  </aug>
  <source>Jmir Infodemiology</source>
  <publisher>JMIR Publications Toronto, Canada</publisher>
  <pubdate>2022</pubdate>
  <volume>2</volume>
  <issue>1</issue>
  <fpage>e32378</fpage>
</bibl>

<bibl id="B18">
  <title><p>The disinformation dozen: An exploratory analysis of covid-19
  disinformation proliferation on twitter</p></title>
  <aug>
    <au><snm>Nogara</snm><fnm>G</fnm></au>
    <au><snm>Vishnuprasad</snm><fnm>PS</fnm></au>
    <au><snm>Cardoso</snm><fnm>F</fnm></au>
    <au><snm>Ayoub</snm><fnm>O</fnm></au>
    <au><snm>Giordano</snm><fnm>S</fnm></au>
    <au><snm>Luceri</snm><fnm>L</fnm></au>
  </aug>
  <source>14th ACM Web Science Conference 2022</source>
  <pubdate>2022</pubdate>
  <fpage>348</fpage>
  <lpage>-358</lpage>
</bibl>

<bibl id="B19">
  <title><p>Evaluating the effectiveness of deplatforming as a moderation
  strategy on Twitter</p></title>
  <aug>
    <au><snm>Jhaver</snm><fnm>S</fnm></au>
    <au><snm>Boylston</snm><fnm>C</fnm></au>
    <au><snm>Yang</snm><fnm>D</fnm></au>
    <au><snm>Bruckman</snm><fnm>A</fnm></au>
  </aug>
  <source>Proceedings of the ACM on Human-Computer Interaction</source>
  <publisher>ACM New York, NY, USA</publisher>
  <pubdate>2021</pubdate>
  <volume>5</volume>
  <issue>CSCW2</issue>
  <fpage>1</fpage>
  <lpage>-30</lpage>
</bibl>

<bibl id="B20">
  <title><p>Comparative analysis of social bots and humans during the COVID-19
  pandemic</p></title>
  <aug>
    <au><snm>Chang</snm><fnm>HCH</fnm></au>
    <au><snm>Ferrara</snm><fnm>E</fnm></au>
  </aug>
  <source>Journal of Computational Social Science</source>
  <publisher>Springer Nature Singapore</publisher>
  <pubdate>2022</pubdate>
  <fpage>1409</fpage>
  <lpage>-1425</lpage>
</bibl>

<bibl id="B21">
  <title><p>Identifying and Characterizing Behavioral Classes of Radicalization
  within the QAnon Conspiracy on Twitter</p></title>
  <aug>
    <au><snm>Wang</snm><fnm>E</fnm></au>
    <au><snm>Luceri</snm><fnm>L</fnm></au>
    <au><snm>Pierri</snm><fnm>F</fnm></au>
    <au><snm>Ferrara</snm><fnm>E</fnm></au>
  </aug>
  <source>17th International Conference on Web and Social Media</source>
  <pubdate>2023</pubdate>
</bibl>

<bibl id="B22">
  <title><p>Social media against society</p></title>
  <aug>
    <au><snm>Luceri</snm><fnm>L</fnm></au>
    <au><snm>Cresci</snm><fnm>S</fnm></au>
    <au><snm>Giordano</snm><fnm>S</fnm></au>
  </aug>
  <source>The Internet and the 2020 Campaign</source>
  <publisher>Rowman \& Littlefield</publisher>
  <pubdate>2021</pubdate>
  <fpage>1</fpage>
</bibl>

<bibl id="B23">
  <title><p>Understanding the effect of deplatforming on social
  networks</p></title>
  <aug>
    <au><snm>Ali</snm><fnm>S</fnm></au>
    <au><snm>Saeed</snm><fnm>MH</fnm></au>
    <au><snm>Aldreabi</snm><fnm>E</fnm></au>
    <au><snm>Blackburn</snm><fnm>J</fnm></au>
    <au><snm>De Cristofaro</snm><fnm>E</fnm></au>
    <au><snm>Zannettou</snm><fnm>S</fnm></au>
    <au><snm>Stringhini</snm><fnm>G</fnm></au>
  </aug>
  <source>13th ACM Web Science Conference 2021</source>
  <pubdate>2021</pubdate>
  <fpage>187</fpage>
  <lpage>-195</lpage>
</bibl>

<bibl id="B24">
  <title><p>The Impact of Twitter Labels on Misinformation Spread and User
  Engagement: Lessons from Trump’s Election Tweets</p></title>
  <aug>
    <au><snm>Papakyriakopoulos</snm><fnm>O</fnm></au>
    <au><snm>Goodman</snm><fnm>E</fnm></au>
  </aug>
  <source>Proceedings of the ACM Web Conference</source>
  <pubdate>2022</pubdate>
  <fpage>2541</fpage>
  <lpage>-2551</lpage>
</bibl>

<bibl id="B25">
  <title><p>" I Won the Election!": An Empirical Analysis of Soft Moderation
  Interventions on Twitter</p></title>
  <aug>
    <au><snm>Zannettou</snm><fnm>S</fnm></au>
  </aug>
  <source>Proceedings of the International AAAI Conference on Web and Social
  Media</source>
  <pubdate>2021</pubdate>
  <volume>15</volume>
  <fpage>865</fpage>
  <lpage>-876</lpage>
</bibl>

<bibl id="B26">
  <title><p>A postmortem of suspended Twitter accounts in the 2016 US
  presidential election</p></title>
  <aug>
    <au><snm>Le</snm><fnm>H</fnm></au>
    <au><snm>Boynton</snm><fnm>GR</fnm></au>
    <au><snm>Shafiq</snm><fnm>Z</fnm></au>
    <au><snm>Srinivasan</snm><fnm>P</fnm></au>
  </aug>
  <source>2019 IEEE/ACM International ASONAM Conference</source>
  <pubdate>2019</pubdate>
  <fpage>258</fpage>
  <lpage>-265</lpage>
</bibl>

<bibl id="B27">
  <title><p>BlackLivesMatter 2020: An Analysis of Deleted and Suspended Users
  in Twitter</p></title>
  <aug>
    <au><snm>Toraman</snm><fnm>C</fnm></au>
    <au><snm>{\c{S}}ahinu{\c{c}}</snm><fnm>F</fnm></au>
    <au><snm>Yilmaz</snm><fnm>EH</fnm></au>
  </aug>
  <source>14th ACM Web Science Conference 2022</source>
  <pubdate>2022</pubdate>
  <fpage>290</fpage>
  <lpage>-295</lpage>
</bibl>

<bibl id="B28">
  <title><p>The role of suspended accounts in political discussion on social
  media: Analysis of the 2017 French, UK and German elections</p></title>
  <aug>
    <au><snm>Maj{\'o} V{\'a}zquez</snm><fnm>S</fnm></au>
    <au><snm>Congosto</snm><fnm>M</fnm></au>
    <au><snm>Nicholls</snm><fnm>T</fnm></au>
    <au><snm>Nielsen</snm><fnm>RK</fnm></au>
  </aug>
  <source>Social Media+ Society</source>
  <publisher>SAGE Publications Sage UK: London, England</publisher>
  <pubdate>2021</pubdate>
</bibl>

<bibl id="B29">
  <title><p>A Large-Scale Longitudinal Multimodal Dataset of State-Backed
  Information Operations on Twitter</p></title>
  <aug>
    <au><snm>Guo</snm><fnm>X</fnm></au>
    <au><snm>Vosoughi</snm><fnm>S</fnm></au>
  </aug>
  <source>Proceedings of the International AAAI Conference on Web and Social
  Media</source>
  <pubdate>2022</pubdate>
</bibl>

<bibl id="B30">
  <title><p>Analyzing the digital traces of political manipulation: The 2016
  Russian interference Twitter campaign</p></title>
  <aug>
    <au><snm>Badawy</snm><fnm>A</fnm></au>
    <au><snm>Ferrara</snm><fnm>E</fnm></au>
    <au><snm>Lerman</snm><fnm>K</fnm></au>
  </aug>
  <source>2018 IEEE/ACM International ASONAM Conference</source>
  <pubdate>2018</pubdate>
  <fpage>258</fpage>
  <lpage>-265</lpage>
</bibl>

<bibl id="B31">
  <title><p>Bots increase exposure to negative and inflammatory content in
  online social systems</p></title>
  <aug>
    <au><snm>Stella</snm><fnm>M</fnm></au>
    <au><snm>Ferrara</snm><fnm>E</fnm></au>
    <au><snm>De Domenico</snm><fnm>M</fnm></au>
  </aug>
  <source>Proceedings of the National Academy of Sciences</source>
  <publisher>National Acad Sciences</publisher>
  <pubdate>2018</pubdate>
  <volume>115</volume>
  <issue>49</issue>
  <fpage>12435</fpage>
  <lpage>-12440</lpage>
</bibl>

<bibl id="B32">
  <title><p>Political polarization drives online conversations about COVID-19
  in the United States</p></title>
  <aug>
    <au><snm>Jiang</snm><fnm>J</fnm></au>
    <au><snm>Chen</snm><fnm>E</fnm></au>
    <au><snm>Yan</snm><fnm>S</fnm></au>
    <au><snm>Lerman</snm><fnm>K</fnm></au>
    <au><snm>Ferrara</snm><fnm>E</fnm></au>
  </aug>
  <source>Human Behavior and Emerging Technologies</source>
  <publisher>Wiley Online Library</publisher>
  <pubdate>2020</pubdate>
  <volume>2</volume>
  <issue>3</issue>
  <fpage>200</fpage>
  <lpage>-211</lpage>
</bibl>

<bibl id="B33">
  <title><p>Suspicious Twitter Activity around the Russian Invasion of
  Ukraine</p></title>
  <aug>
    <au><cnm>{IU Observatory on Social Media}</cnm></au>
  </aug>
  <pubdate>2022</pubdate>
</bibl>

<bibl id="B34">
  <title><p>Analysis of Twitter accounts created around the invasion of
  Ukraine</p></title>
  <aug>
    <au><cnm>{IU Observatory on Social Media}</cnm></au>
  </aug>
  <pubdate>2022</pubdate>
</bibl>

<bibl id="B35">
  <title><p>Disinformation and social bot operations in the run up to the 2017
  French presidential election</p></title>
  <aug>
    <au><snm>Ferrara</snm><fnm>E</fnm></au>
  </aug>
  <source>First Monday</source>
  <pubdate>2017</pubdate>
  <volume>22</volume>
  <issue>8</issue>
</bibl>

<bibl id="B36">
  <title><p>Detecting spammers on social networks</p></title>
  <aug>
    <au><snm>Stringhini</snm><fnm>G</fnm></au>
    <au><snm>Kruegel</snm><fnm>C</fnm></au>
    <au><snm>Vigna</snm><fnm>G</fnm></au>
  </aug>
  <source>Proceedings of the 26th annual computer security applications
  conference</source>
  <pubdate>2010</pubdate>
  <fpage>1</fpage>
  <lpage>-9</lpage>
</bibl>

<bibl id="B37">
  <title><p>Analyzing spammers' social networks for fun and profit: a case
  study of cyber criminal ecosystem on twitter</p></title>
  <aug>
    <au><snm>Yang</snm><fnm>C</fnm></au>
    <au><snm>Harkreader</snm><fnm>R</fnm></au>
    <au><snm>Zhang</snm><fnm>J</fnm></au>
    <au><snm>Shin</snm><fnm>S</fnm></au>
    <au><snm>Gu</snm><fnm>G</fnm></au>
  </aug>
  <source>Proceedings of the 21st international conference on World Wide
  Web</source>
  <pubdate>2012</pubdate>
  <fpage>71</fpage>
  <lpage>-80</lpage>
</bibl>

<bibl id="B38">
  <title><p>The history of digital spam</p></title>
  <aug>
    <au><snm>Ferrara</snm><fnm>E</fnm></au>
  </aug>
  <source>Communications of the ACM</source>
  <pubdate>2019</pubdate>
  <volume>62</volume>
  <issue>8</issue>
  <fpage>82</fpage>
  <lpage>-91</lpage>
</bibl>

<bibl id="B39">
  <title><p>Twitter spam and false accounts prevalence, detection, and
  characterization: A survey</p></title>
  <aug>
    <au><snm>Ferrara</snm><fnm>E</fnm></au>
  </aug>
  <source>First Monday</source>
  <pubdate>2022</pubdate>
  <volume>27</volume>
  <issue>12</issue>
</bibl>

<bibl id="B40">
  <title><p>Detecting troll behavior via inverse reinforcement learning: A case
  study of russian trolls in the 2016 us election</p></title>
  <aug>
    <au><snm>Luceri</snm><fnm>L</fnm></au>
    <au><snm>Giordano</snm><fnm>S</fnm></au>
    <au><snm>Ferrara</snm><fnm>E</fnm></au>
  </aug>
  <source>Proceedings of the international AAAI conference on web and social
  media</source>
  <pubdate>2020</pubdate>
  <volume>14</volume>
  <fpage>417</fpage>
  <lpage>-427</lpage>
</bibl>

<bibl id="B41">
  <title><p>Investigating the difference between trolls, social bots, and
  humans on Twitter</p></title>
  <aug>
    <au><snm>Mazza</snm><fnm>M</fnm></au>
    <au><snm>Avvenuti</snm><fnm>M</fnm></au>
    <au><snm>Cresci</snm><fnm>S</fnm></au>
    <au><snm>Tesconi</snm><fnm>M</fnm></au>
  </aug>
  <source>Computer Communications</source>
  <publisher>Elsevier</publisher>
  <pubdate>2022</pubdate>
  <volume>196</volume>
  <fpage>23</fpage>
  <lpage>-36</lpage>
</bibl>

<bibl id="B42">
  <title><p>Linguistic cues to deception: Identifying political trolls on
  social media</p></title>
  <aug>
    <au><snm>Addawood</snm><fnm>A</fnm></au>
    <au><snm>Badawy</snm><fnm>A</fnm></au>
    <au><snm>Lerman</snm><fnm>K</fnm></au>
    <au><snm>Ferrara</snm><fnm>E</fnm></au>
  </aug>
  <source>Proceedings of the international AAAI conference on web and social
  media</source>
  <pubdate>2019</pubdate>
  <volume>13</volume>
  <fpage>15</fpage>
  <lpage>-25</lpage>
</bibl>

<bibl id="B43">
  <title><p>On Twitter purge: a retrospective analysis of suspended
  users</p></title>
  <aug>
    <au><snm>Chowdhury</snm><fnm>FA</fnm></au>
    <au><snm>Allen</snm><fnm>L</fnm></au>
    <au><snm>Yousuf</snm><fnm>M</fnm></au>
    <au><snm>Mueen</snm><fnm>A</fnm></au>
  </aug>
  <source>Companion proceedings of the web conference</source>
  <pubdate>2020</pubdate>
  <fpage>371</fpage>
  <lpage>-378</lpage>
</bibl>

<bibl id="B44">
  <title><p>Examining factors associated with twitter account suspension
  following the 2020 us presidential election</p></title>
  <aug>
    <au><snm>Chowdhury</snm><fnm>FA</fnm></au>
    <au><snm>Saha</snm><fnm>D</fnm></au>
    <au><snm>Hasan</snm><fnm>MR</fnm></au>
    <au><snm>Saha</snm><fnm>K</fnm></au>
    <au><snm>Mueen</snm><fnm>A</fnm></au>
  </aug>
  <source>Proceedings of the 2021 IEEE/ACM International Conference on Advances
  in Social Networks Analysis and Mining</source>
  <pubdate>2021</pubdate>
  <fpage>607</fpage>
  <lpage>-612</lpage>
</bibl>

<bibl id="B45">
  <title><p>Textual Analysis and Timely Detection of Suspended Social Media
  Accounts.</p></title>
  <aug>
    <au><snm>Seyler</snm><fnm>D</fnm></au>
    <au><snm>Tan</snm><fnm>S</fnm></au>
    <au><snm>Li</snm><fnm>D</fnm></au>
    <au><snm>Zhang</snm><fnm>J</fnm></au>
    <au><snm>Li</snm><fnm>P</fnm></au>
  </aug>
  <source>ICWSM</source>
  <pubdate>2021</pubdate>
  <fpage>644</fpage>
  <lpage>-655</lpage>
</bibl>

<bibl id="B46">
  <title><p>Characterizing the 2022 Russo-Ukrainian Conflict Through the Lenses
  of Aspect-Based Sentiment Analysis: Dataset, Methodology, and Preliminary
  Findings</p></title>
  <aug>
    <au><snm>Caprolu</snm><fnm>M</fnm></au>
    <au><snm>Sadighian</snm><fnm>A</fnm></au>
    <au><snm>Di Pietro</snm><fnm>R</fnm></au>
  </aug>
  <source>arXiv:2208.04903</source>
  <pubdate>2022</pubdate>
</bibl>

<bibl id="B47">
  <title><p>VoynaSlov: A Data Set of Russian Social Media Activity during the
  2022 Ukraine-Russia War</p></title>
  <aug>
    <au><snm>Park</snm><fnm>CY</fnm></au>
    <au><snm>Mendelsohn</snm><fnm>J</fnm></au>
    <au><snm>Field</snm><fnm>A</fnm></au>
    <au><snm>Tsvetkov</snm><fnm>Y</fnm></au>
  </aug>
  <source>arXiv:2205.12382</source>
  <pubdate>2022</pubdate>
</bibl>

<bibl id="B48">
  <title><p>Happenstance: Utilizing Semantic Search to Track Russian State
  Media Narratives about the Russo-Ukrainian War On Reddit</p></title>
  <aug>
    <au><snm>Hanley</snm><fnm>HW</fnm></au>
    <au><snm>Kumar</snm><fnm>D</fnm></au>
    <au><snm>Durumeric</snm><fnm>Z</fnm></au>
  </aug>
  <source>arXiv:2205.14484</source>
  <pubdate>2022</pubdate>
</bibl>

<bibl id="B49">
  <title><p>" A Special Operation": A Quantitative Approach to Dissecting and
  Comparing Different Media Ecosystems' Coverage of the Russo-Ukrainian
  War</p></title>
  <aug>
    <au><snm>Hanley</snm><fnm>HW</fnm></au>
    <au><snm>Kumar</snm><fnm>D</fnm></au>
    <au><snm>Durumeric</snm><fnm>Z</fnm></au>
  </aug>
  <source>arXiv:2210.03016</source>
  <pubdate>2022</pubdate>
</bibl>

<bibl id="B50">
  <title><p>Russian propaganda on social media during the 2022 invasion of
  Ukraine</p></title>
  <aug>
    <au><snm>Geissler</snm><fnm>D</fnm></au>
    <au><snm>B{\"a}r</snm><fnm>D</fnm></au>
    <au><snm>Pr{\"o}llochs</snm><fnm>N</fnm></au>
    <au><snm>Feuerriegel</snm><fnm>S</fnm></au>
  </aug>
  <source>arXiv:2211.04154</source>
  <pubdate>2022</pubdate>
</bibl>

<bibl id="B51">
  <title><p>Propaganda and Misinformation on Facebook and Twitter during the
  Russian Invasion of Ukraine</p></title>
  <aug>
    <au><snm>Pierri</snm><fnm>F</fnm></au>
    <au><snm>Luceri</snm><fnm>L</fnm></au>
    <au><snm>Jindal</snm><fnm>N</fnm></au>
    <au><snm>Ferrara</snm><fnm>E</fnm></au>
  </aug>
  <source>WebSci’23 -- 15th ACM Web Science Conference</source>
  <pubdate>2023</pubdate>
</bibl>

<bibl id="B52">
  <title><p>The limited reach of fake news on Twitter during 2019 European
  elections</p></title>
  <aug>
    <au><snm>Cinelli</snm><fnm>M</fnm></au>
    <au><snm>Cresci</snm><fnm>S</fnm></au>
    <au><snm>Galeazzi</snm><fnm>A</fnm></au>
    <au><snm>Quattrociocchi</snm><fnm>W</fnm></au>
    <au><snm>Tesconi</snm><fnm>M</fnm></au>
  </aug>
  <source>PloS one</source>
  <publisher>Public Library of Science San Francisco, CA USA</publisher>
  <pubdate>2020</pubdate>
  <volume>15</volume>
  <issue>6</issue>
  <fpage>e0234689</fpage>
</bibl>

<bibl id="B53">
  <title><p>Fake news on Twitter during the 2016 US presidential
  election</p></title>
  <aug>
    <au><snm>Grinberg</snm><fnm>N</fnm></au>
    <au><snm>Joseph</snm><fnm>K</fnm></au>
    <au><snm>Friedland</snm><fnm>L</fnm></au>
    <au><snm>Swire Thompson</snm><fnm>B</fnm></au>
    <au><snm>Lazer</snm><fnm>D</fnm></au>
  </aug>
  <source>Science</source>
  <publisher>American Association for the Advancement of Science</publisher>
  <pubdate>2019</pubdate>
  <volume>363</volume>
  <issue>6425</issue>
  <fpage>374</fpage>
  <lpage>-378</lpage>
</bibl>

<bibl id="B54">
  <title><p>Bots, elections, and social media: a brief overview</p></title>
  <aug>
    <au><snm>Ferrara</snm><fnm>E</fnm></au>
  </aug>
  <source>Disinformation, Misinformation, and Fake News in Social
  Media</source>
  <publisher>Springer</publisher>
  <pubdate>2020</pubdate>
  <fpage>95</fpage>
  <lpage>-114</lpage>
</bibl>

<bibl id="B55">
  <title><p>Retweet-BERT: Political Leaning Detection Using Language Features
  and Information Diffusion on Social Networks</p></title>
  <aug>
    <au><snm>Jiang</snm><fnm>J</fnm></au>
    <au><snm>Ren</snm><fnm>X</fnm></au>
    <au><snm>Ferrara</snm><fnm>E</fnm></au>
  </aug>
  <source>17th International AAAI Conference on Web and Social Media</source>
  <pubdate>2023</pubdate>
</bibl>

<bibl id="B56">
  <title><p>Political Communities on Twitter: Case Study of the 2022 French
  Presidential Election</p></title>
  <aug>
    <au><snm>Abdine</snm><fnm>H</fnm></au>
    <au><snm>Guo</snm><fnm>Y</fnm></au>
    <au><snm>Rennard</snm><fnm>V</fnm></au>
    <au><snm>Vazirgiannis</snm><fnm>M</fnm></au>
  </aug>
  <source>arXiv:2204.07436</source>
  <pubdate>2022</pubdate>
</bibl>

<bibl id="B57">
  <title><p>Tweets in Time of Conflict: A Public Dataset Tracking the Twitter
  Discourse on the War Between Ukraine and Russia</p></title>
  <aug>
    <au><snm>Chen</snm><fnm>E</fnm></au>
    <au><snm>Ferrara</snm><fnm>E</fnm></au>
  </aug>
  <source>17th International AAAI Conference on Web and Social Media
  (ICWSM'23)</source>
  <pubdate>2023</pubdate>
</bibl>

<bibl id="B58">
  <title><p>Ukraine Twitter data</p></title>
  <aug>
    <au><snm>Munch</snm><fnm>FV</fnm></au>
    <au><snm>Kessling</snm><fnm>P</fnm></au>
  </aug>
  <publisher>OSF</publisher>
  <pubdate>2022</pubdate>
  <url>osf.io/rtqxn</url>
</bibl>

<bibl id="B59">
  <title><p>CoVaxxy: A global collection of English Twitter posts about
  COVID-19 vaccines</p></title>
  <aug>
    <au><snm>DeVerna</snm><fnm>M</fnm></au>
    <au><snm>Pierri</snm><fnm>F</fnm></au>
    <au><snm>Truong</snm><fnm>B</fnm></au>
    <au><snm>Bollenbacher</snm><fnm>J</fnm></au>
    <au><snm>Axelrod</snm><fnm>D</fnm></au>
    <au><snm>Loynes</snm><fnm>N</fnm></au>
    <au><snm>Torres Lugo</snm><fnm>C</fnm></au>
    <au><snm>Yang</snm><fnm>KC</fnm></au>
    <au><snm>Menczer</snm><fnm>F</fnm></au>
    <au><snm>Bryden</snm><fnm>J</fnm></au>
  </aug>
  <source>Proceedings of the International AAAI Conference on Web and Social
  Media</source>
  <pubdate>2021</pubdate>
</bibl>

<bibl id="B60">
  <title><p>Is the sample good enough? comparing data from twitter's streaming
  api with twitter's firehose</p></title>
  <aug>
    <au><snm>Morstatter</snm><fnm>F</fnm></au>
    <au><snm>Pfeffer</snm><fnm>J</fnm></au>
    <au><snm>Liu</snm><fnm>H</fnm></au>
    <au><snm>Carley</snm><fnm>K</fnm></au>
  </aug>
  <source>Proceedings of the international AAAI conference on web and social
  media</source>
  <pubdate>2013</pubdate>
  <volume>7</volume>
  <issue>1</issue>
  <fpage>400</fpage>
  <lpage>-408</lpage>
</bibl>

<bibl id="B61">
  <title><p>Detecting Harmful Content on Online Platforms: What Platforms Need
  vs. Where Research Efforts Go</p></title>
  <aug>
    <au><snm>Arora</snm><fnm>A</fnm></au>
    <au><snm>Nakov</snm><fnm>P</fnm></au>
    <au><snm>Hardalov</snm><fnm>M</fnm></au>
    <au><snm>Sarwar</snm><fnm>SM</fnm></au>
    <au><snm>Nayak</snm><fnm>V</fnm></au>
    <au><snm>Dinkov</snm><fnm>Y</fnm></au>
    <au><snm>Zlatkova</snm><fnm>D</fnm></au>
    <au><snm>Dent</snm><fnm>K</fnm></au>
    <au><snm>Bhatawdekar</snm><fnm>A</fnm></au>
    <au><snm>Bouchard</snm><fnm>G</fnm></au>
    <au><cnm>others</cnm></au>
  </aug>
  <source>ACM Computing Surveys</source>
  <publisher>ACM New York, NY</publisher>
  <pubdate>2023</pubdate>
</bibl>

<bibl id="B62">
  <title><p>Manipulating Twitter through Deletions</p></title>
  <aug>
    <au><snm>Torres Lugo</snm><fnm>C</fnm></au>
    <au><snm>Pote</snm><fnm>M</fnm></au>
    <au><snm>Nwala</snm><fnm>AC</fnm></au>
    <au><snm>Menczer</snm><fnm>F</fnm></au>
  </aug>
  <source>Proceedings of the International AAAI Conference on Web and Social
  Media</source>
  <pubdate>2022</pubdate>
  <volume>16</volume>
  <fpage>1029</fpage>
  <lpage>-1039</lpage>
</bibl>

<bibl id="B63">
  <title><p>Red bots do it better: Comparative analysis of social bot partisan
  behavior</p></title>
  <aug>
    <au><snm>Luceri</snm><fnm>L</fnm></au>
    <au><snm>Deb</snm><fnm>A</fnm></au>
    <au><snm>Badawy</snm><fnm>A</fnm></au>
    <au><snm>Ferrara</snm><fnm>E</fnm></au>
  </aug>
  <source>Companion proceedings of the 2019 world wide web conference</source>
  <pubdate>2019</pubdate>
  <fpage>1007</fpage>
  <lpage>-1012</lpage>
</bibl>

<bibl id="B64">
  <title><p>Online Networks of Support in Distressed Environments: Solidarity
  and Mobilization during the Russian Invasion of Ukraine</p></title>
  <aug>
    <au><snm>Ye</snm><fnm>J</fnm></au>
    <au><snm>Jindal</snm><fnm>N</fnm></au>
    <au><snm>Pierri</snm><fnm>F</fnm></au>
    <au><snm>Luceri</snm><fnm>L</fnm></au>
  </aug>
  <source>Companion proceedings of ICWSM 2023</source>
  <pubdate>2023</pubdate>
</bibl>

<bibl id="B65">
  <title><p>Charting the landscape of online cryptocurrency
  manipulation</p></title>
  <aug>
    <au><snm>Nizzoli</snm><fnm>L</fnm></au>
    <au><snm>Tardelli</snm><fnm>S</fnm></au>
    <au><snm>Avvenuti</snm><fnm>M</fnm></au>
    <au><snm>Cresci</snm><fnm>S</fnm></au>
    <au><snm>Tesconi</snm><fnm>M</fnm></au>
    <au><snm>Ferrara</snm><fnm>E</fnm></au>
  </aug>
  <source>IEEE Access</source>
  <publisher>IEEE</publisher>
  <pubdate>2020</pubdate>
  <volume>8</volume>
  <fpage>113230</fpage>
  <lpage>-113245</lpage>
</bibl>

<bibl id="B66">
  <title><p>Detecting cryptocurrency pump-and-dump frauds using market and
  social signals</p></title>
  <aug>
    <au><snm>Nghiem</snm><fnm>H</fnm></au>
    <au><snm>Muric</snm><fnm>G</fnm></au>
    <au><snm>Morstatter</snm><fnm>F</fnm></au>
    <au><snm>Ferrara</snm><fnm>E</fnm></au>
  </aug>
  <source>Expert Systems with Applications</source>
  <publisher>Pergamon</publisher>
  <pubdate>2021</pubdate>
  <volume>182</volume>
  <fpage>115284</fpage>
</bibl>

<bibl id="B67">
  <title><p>Large scale crowdsourcing and characterization of twitter abusive
  behavior</p></title>
  <aug>
    <au><snm>Founta</snm><fnm>AM</fnm></au>
    <au><snm>Djouvas</snm><fnm>C</fnm></au>
    <au><snm>Chatzakou</snm><fnm>D</fnm></au>
    <au><snm>Leontiadis</snm><fnm>I</fnm></au>
    <au><snm>Blackburn</snm><fnm>J</fnm></au>
    <au><snm>Stringhini</snm><fnm>G</fnm></au>
    <au><snm>Vakali</snm><fnm>A</fnm></au>
    <au><snm>Sirivianos</snm><fnm>M</fnm></au>
    <au><snm>Kourtellis</snm><fnm>N</fnm></au>
  </aug>
  <source>Twelfth International AAAI Conference on Web and Social
  Media</source>
  <pubdate>2018</pubdate>
</bibl>

<bibl id="B68">
  <title><p>Botometer 101: Social bot practicum for computational social
  scientists</p></title>
  <aug>
    <au><snm>Yang</snm><fnm>KC</fnm></au>
    <au><snm>Ferrara</snm><fnm>E</fnm></au>
    <au><snm>Menczer</snm><fnm>F</fnm></au>
  </aug>
  <source>Journal of computational social science</source>
  <pubdate>2022</pubdate>
  <volume>5</volume>
  <fpage>1511</fpage>
  <lpage>-1528</lpage>
</bibl>

<bibl id="B69">
  <title><p>What types of COVID-19 conspiracies are populated by Twitter
  bots?</p></title>
  <aug>
    <au><snm>Ferrara</snm><fnm>E</fnm></au>
  </aug>
  <source>arXiv preprint arXiv:2004.09531</source>
  <pubdate>2020</pubdate>
</bibl>

<bibl id="B70">
  <title><p>Down the bot hole: Actionable insights from a one-year analysis of
  bot activity on Twitter</p></title>
  <aug>
    <au><snm>Luceri</snm><fnm>L</fnm></au>
    <au><snm>Cardoso</snm><fnm>F</fnm></au>
    <au><snm>Giordano</snm><fnm>S</fnm></au>
  </aug>
  <source>First Monday</source>
  <pubdate>2021</pubdate>
</bibl>

</refgrp>
} % end of \BMCxmlcomment
% for author-year bibliography (bmc-mathphys or spbasic)
% a) write to bib file (bmc-mathphys only)
% @settings{label, options=''nameyear''}

% or include bibliography directly:
% \begin{thebibliography}
% \bibitem{b1}
% \end{thebibliography}

%%%%%%%%%%%%%%%%%%%%%%%%%%%%%%%%%%%
%%                               %%
%% Figures                       %%
%%                               %%
%% NB: this is for captions and  %%
%% Titles. All graphics must be  %%
%% submitted separately and NOT  %%
%% included in the Tex document  %%
%%                               %%
%%%%%%%%%%%%%%%%%%%%%%%%%%%%%%%%%%%

%%
%% Do not use \listoffigures as most will included as separate files

\section*{Figure legends}
\begin{figure}[!t]
    \centering
    \includegraphics[width=\linewidth]{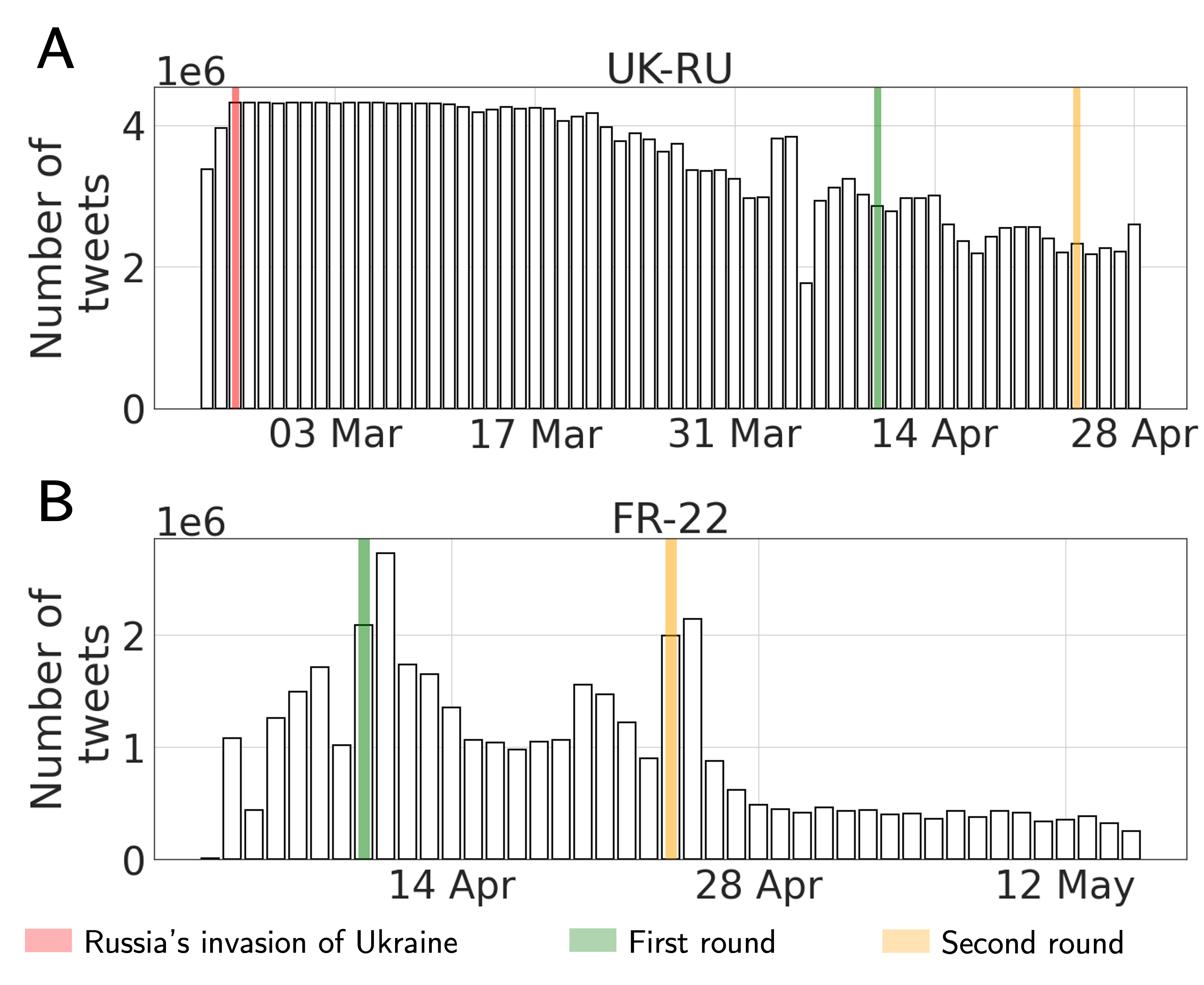}
    \caption{Time series of the daily number of tweets collected in \texttt{UK-RU} \textbf{(left)} and \texttt{FR-22} \textbf{(right)} datasets. Notice the drop in volume on April 7th for \texttt{UK-RU}, which corresponds to a network malfunctioning failure lasting a few hours. N.B: the x-axes are not aligned in the two subplots.}
    \label{fig:tweets-ts}
\end{figure}

\begin{figure}[!t]
    \centering
    \includegraphics[width=\linewidth]{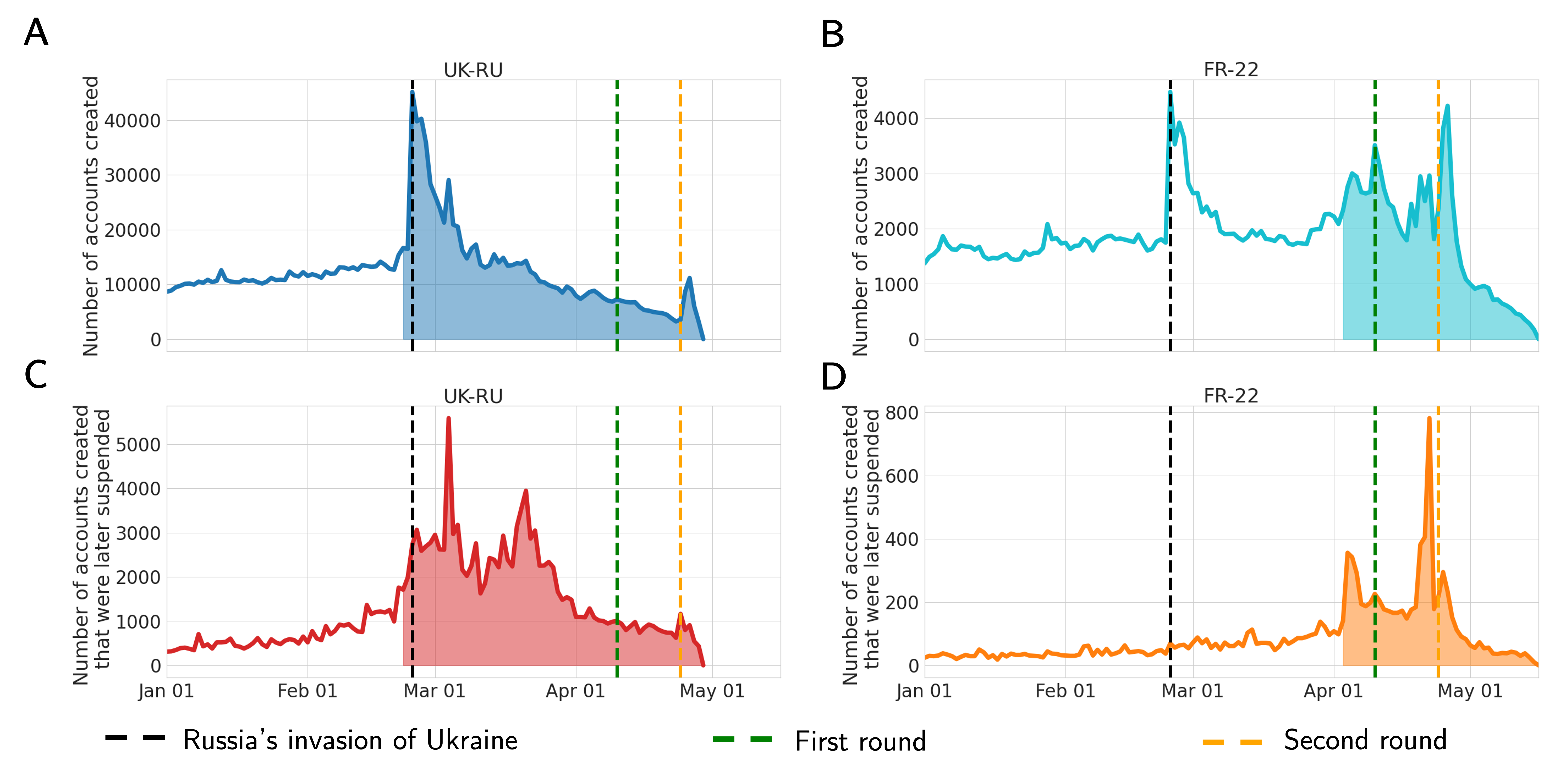}
    \caption{Time series of the daily number of accounts created in \texttt{UK-RU} \textbf{(A)} and \texttt{FR-22} \textbf{(B)} since January 1st to May 15th 2022. Time series of the daily number of accounts created and that were later suspended (as of May 23rd) in \texttt{UK-RU} \textbf{(C)} and \texttt{FR-22} \textbf{(D)}. Colored areas indicate the actual collection period in each dataset. The creation date of accounts is available in the Twitter user object provided by the API.}
    \label{fig:ts-created-suspended}
\end{figure}

\begin{figure}[!t]
    \centering
    \includegraphics[width=\linewidth]{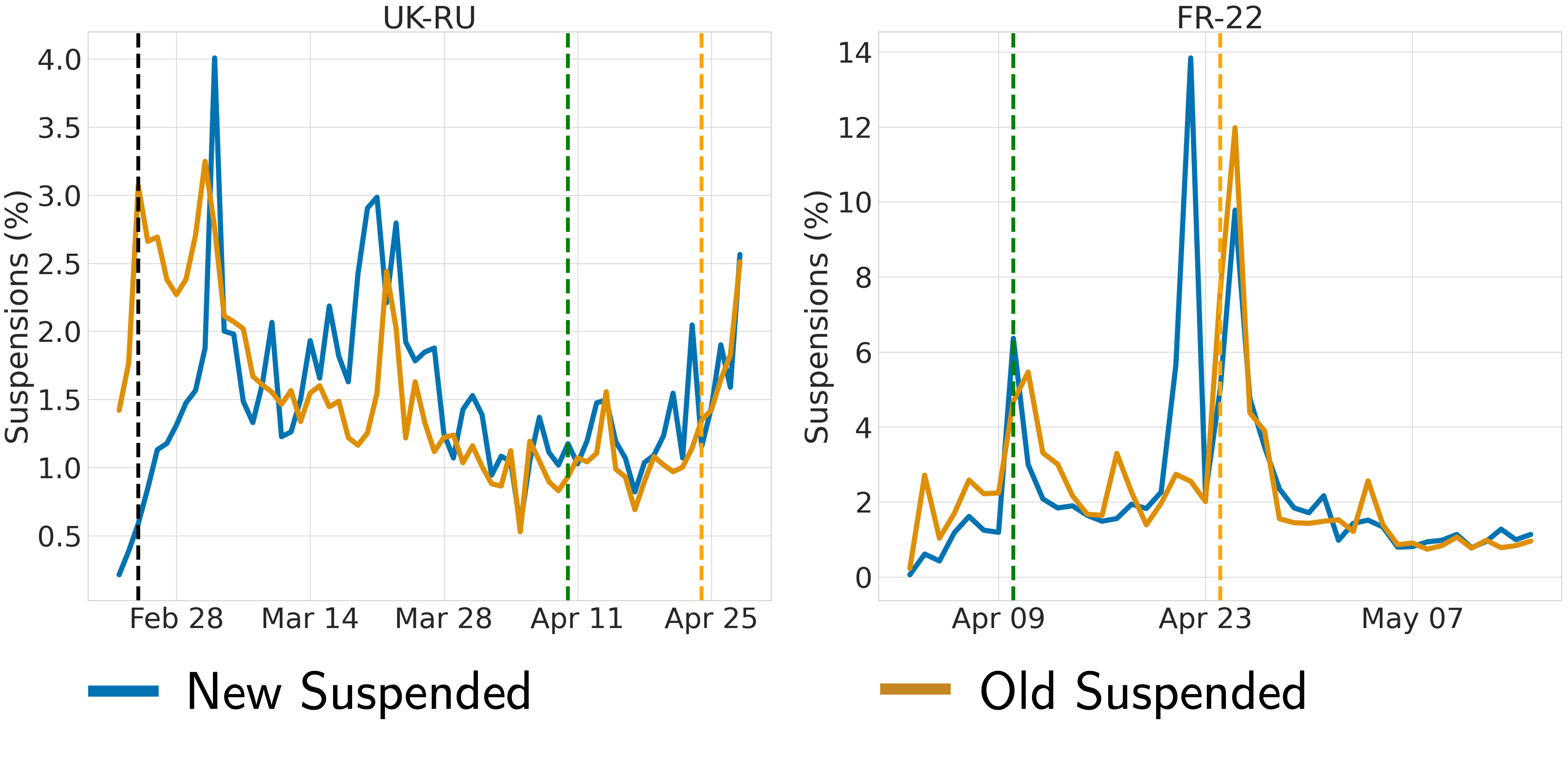}
    \caption{Time series of the daily proportion of accounts that were suspended in \texttt{UK-RU} \textbf{(left)} and \texttt{FR-22} \textbf{(right)} out of all suspended accounts in each dataset, separating accounts created during the collection period (New) from those already existing (Old). Vertical lines indicate the invasion of Ukraine (black) and the two rounds of elections (green and light orange). N.B: the x-axes are not aligned in the two subplots.}
    \label{fig:ts-suspension}
\end{figure}

\begin{figure}[!t]
    \centering
    \includegraphics[width=\linewidth]{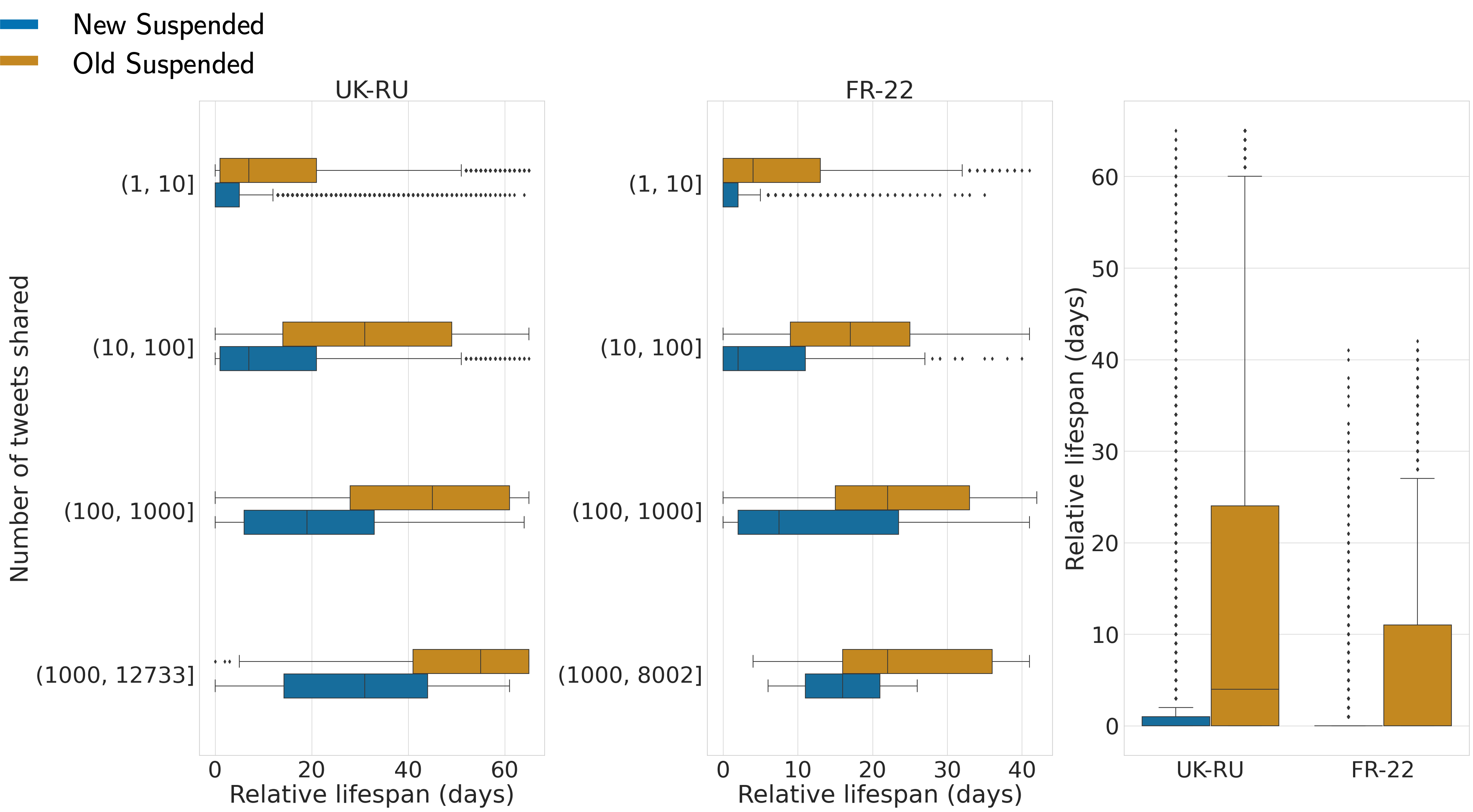}
    \caption{Distribution of relative lifespan matching users on the number of tweets shared in the two datasets, for \texttt{UK-RU} \textbf{(left)} and \texttt{FR-22} \textbf{(center)}. For \texttt{UK-RU} the median values are 0 days and 4 days, respectively for \emph{New} and \emph{Old} suspended users; for \texttt{FR-22} it is 0 days for both \emph{New} and \emph{Old} suspended users. Distribution of the relative lifespan of \emph{New} and \emph{Old} suspended users in both datasets \textbf{(right)}. All distributions of the two classes of users are statistically different according to two-sided Mann-Whitney tests ($p < 0.001$).}
    \label{fig:relative-suspension}
\end{figure}

\begin{figure}[!t]
    \centering
    \includegraphics[width=\linewidth]{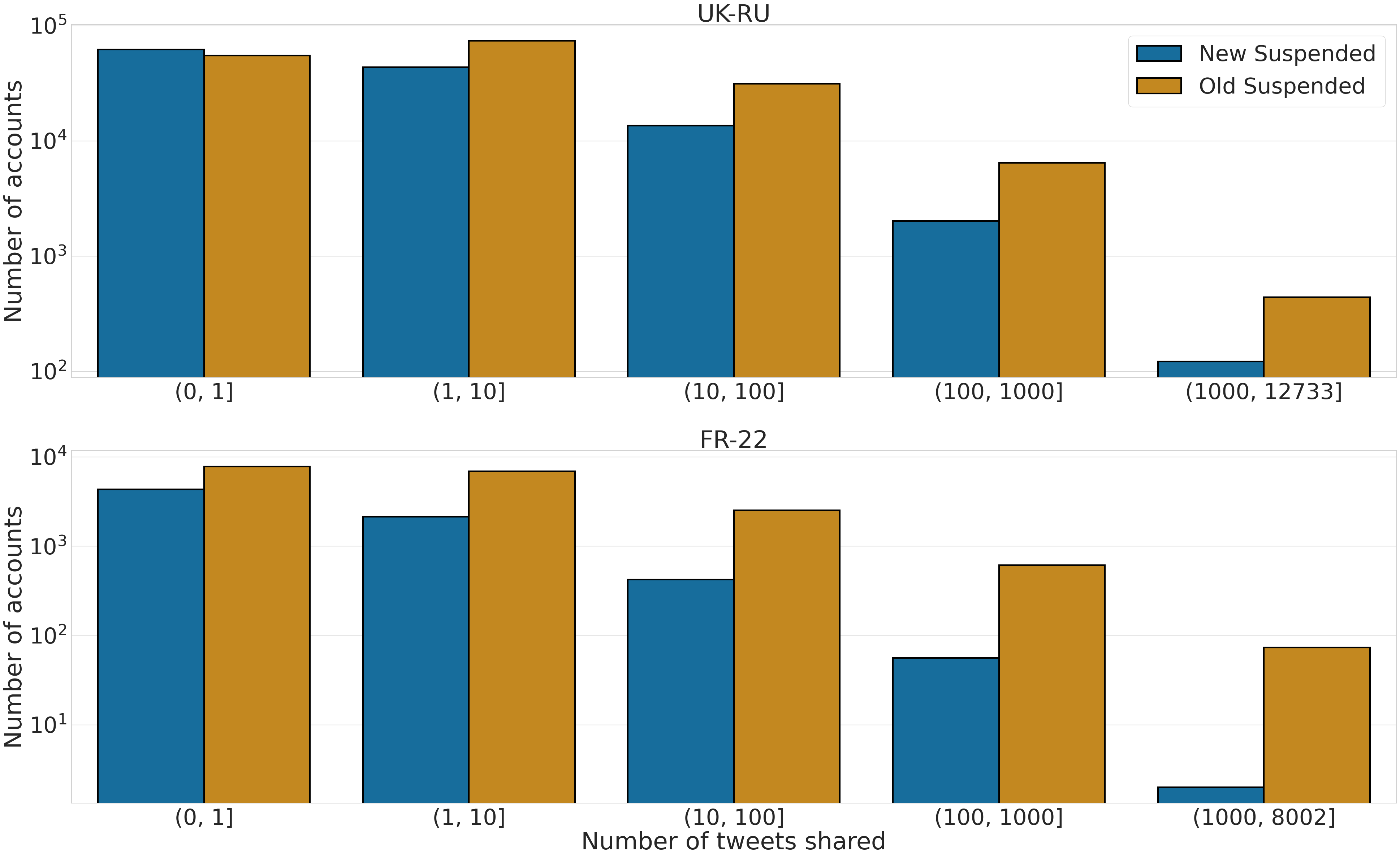}
    \caption{Distribution of the number of New and Old Suspended users for \texttt{UK-RU} \textbf{(top)} and \texttt{FR-22} \textbf{(bottom)}, binned by the number of tweets shared.}
    \label{fig:suspended-bins}
\end{figure}

\begin{figure}[!t]
    \centering
    \includegraphics[width=\linewidth]{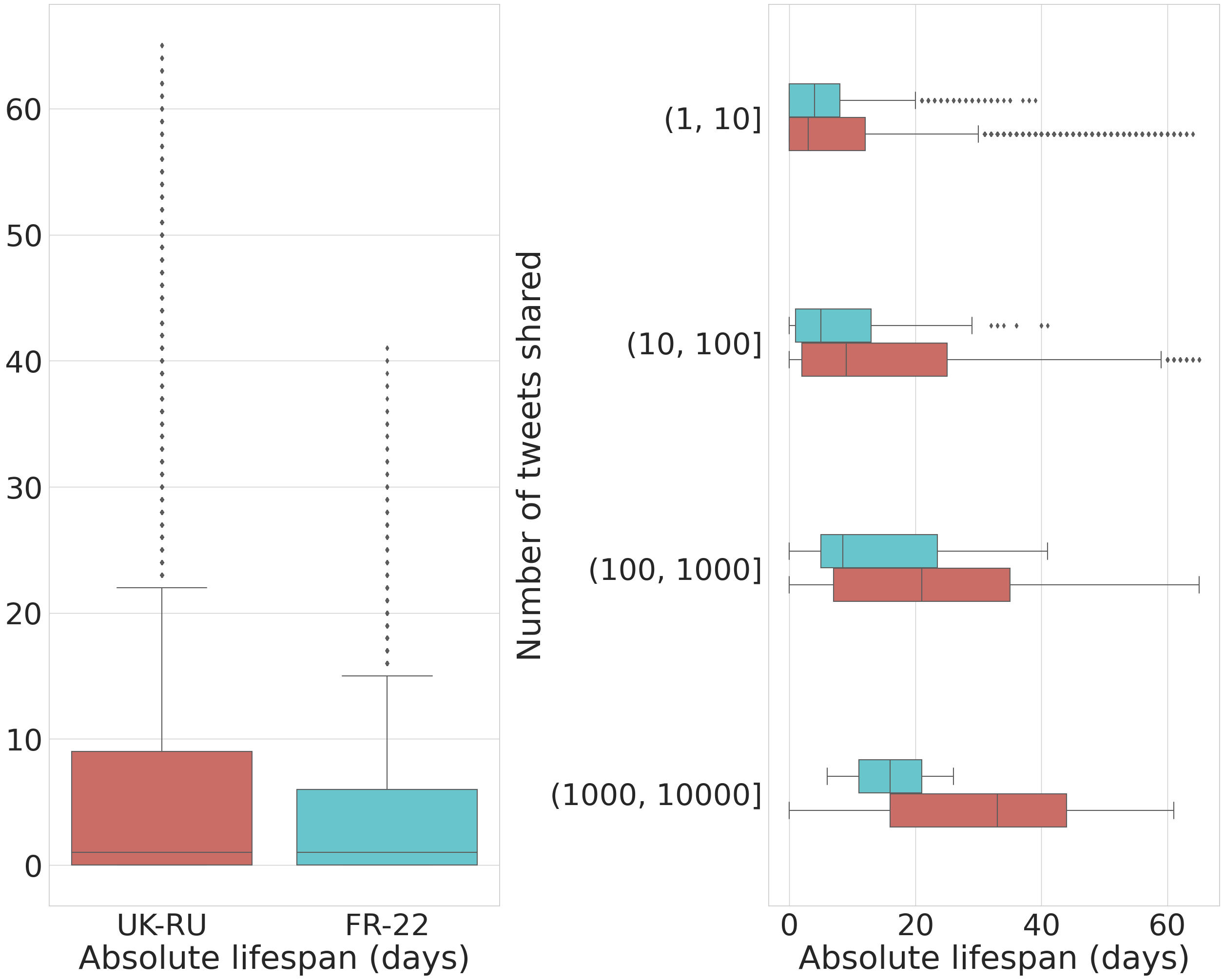}
    \caption{Distribution of the absolute lifespan of \emph{New} suspended users \textbf{(left)} in both datasets (median value is 1 day for both datasets). Same distribution but matching accounts on the number of tweets shared \textbf{(right)}.}
    \label{fig:absolute-suspension}
\end{figure}

\begin{figure}[!t]
    \centering
    \includegraphics[width=\linewidth]{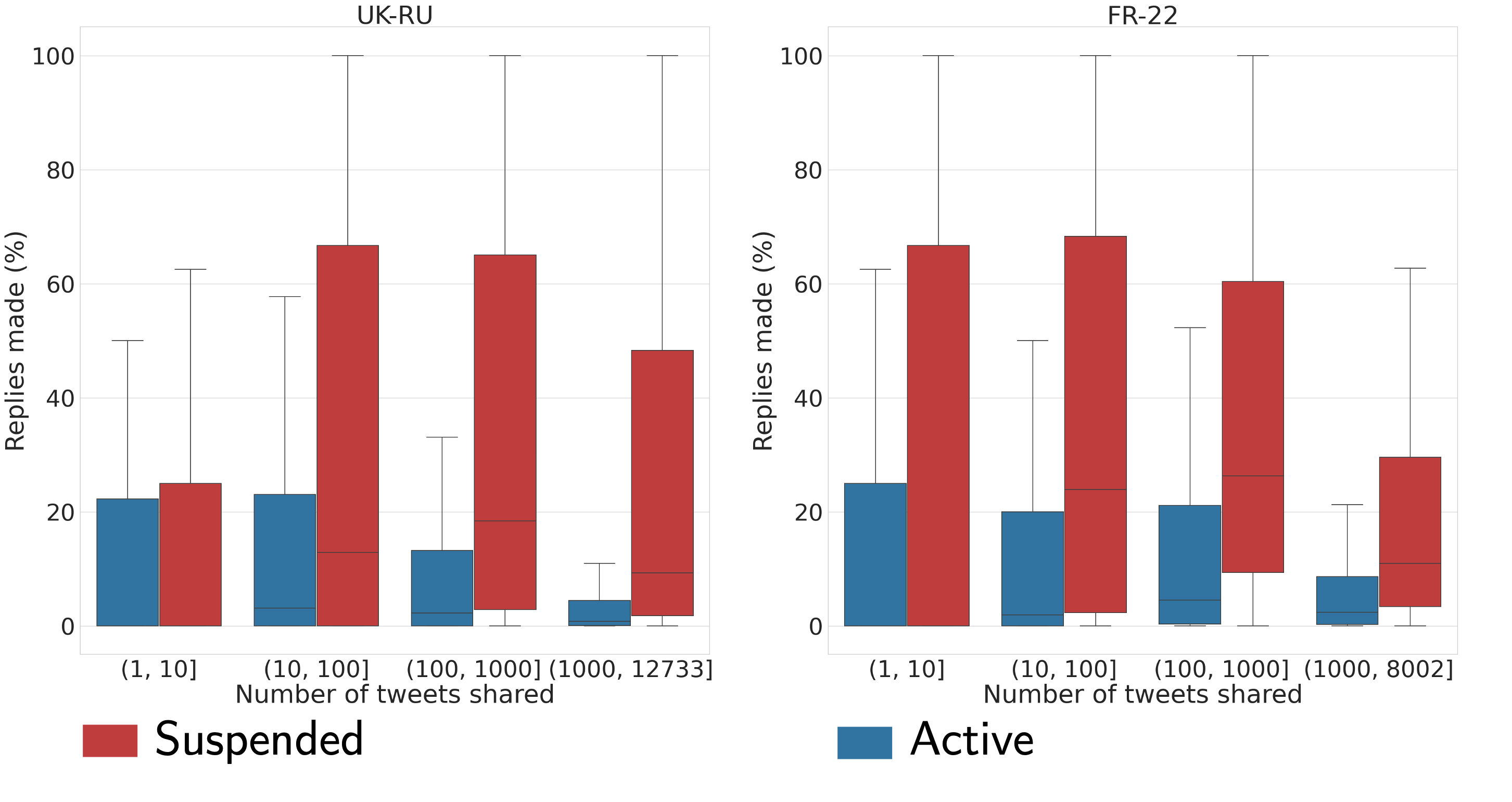}
    \caption{Distribution of the proportion of tweets that are replies for \emph{Active} and \emph{Suspended} accounts in \texttt{UK-RU} \textbf{(left)} and \texttt{FR-22} \textbf{(right)}. Boxplots do not show outliers.}
    \label{fig:boxplot-replies}
\end{figure}

\begin{figure}[!t]
    \centering
    \includegraphics[width=\linewidth]{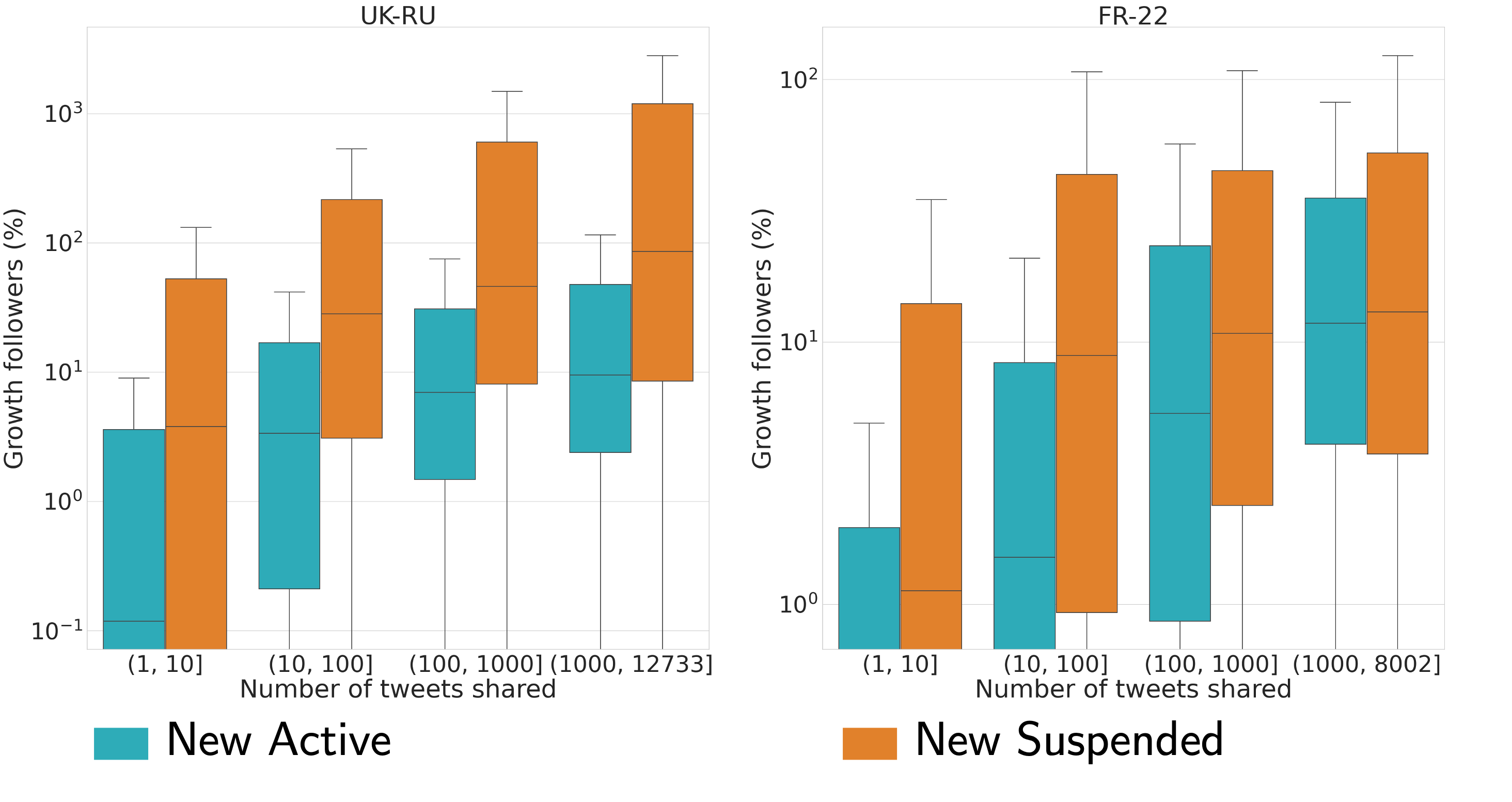}
    \caption{Distribution of the proportion of growth in followers (in percentage) for \emph{New Active} and \emph{New Suspended} users in \texttt{UK-RU} \textbf{(left)} and \texttt{FR-22} \textbf{(right)}. Boxplots do not show outliers. The y-scale is logarithmic.}
    \label{fig:growth-followers}
\end{figure}

\begin{figure}[!t]
    \centering
    \includegraphics[width=\linewidth]{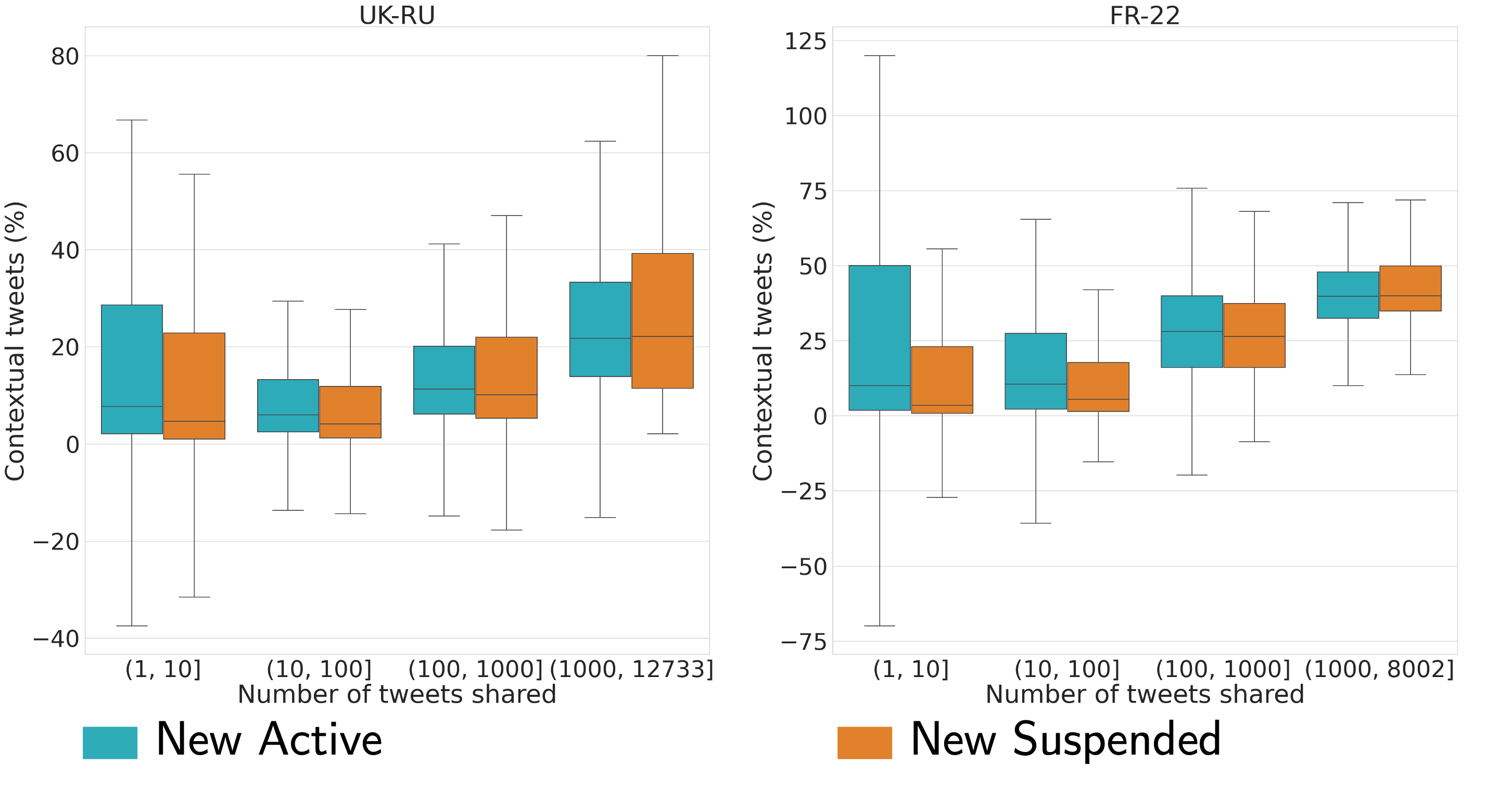}
    \caption{Distribution of the proportion of contextual tweets for \emph{New Active} and \emph{New Suspended} users in \texttt{UK-RU} \textbf{(left)} and \texttt{FR-22} \textbf{(right)}. Boxplots do not show outliers. Negative values are due to accounts that deleted tweets during the period of observation, thus the difference between the final and the initial number of statutes is negative.}
    \label{fig:contextual-tweets-boxplots}
\end{figure}

\begin{figure}[!t]
    \centering
    \includegraphics[width=\linewidth]{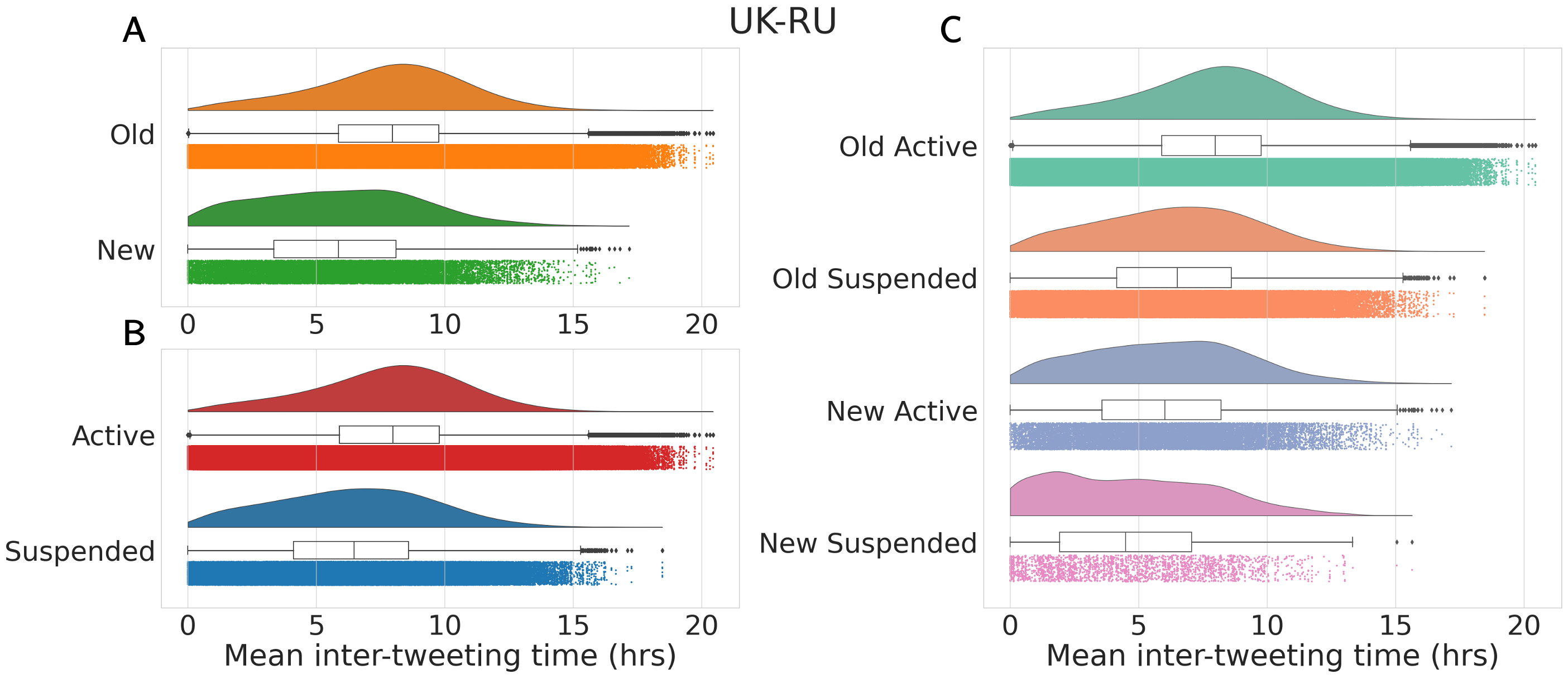}
    \caption{Distribution of the mean inter-tweeting time (in hours) for \emph{Active} versus \emph{Suspended} accounts \textbf{(A)}, \emph{Old} versus \emph{New} users \textbf{(B)} and all four classes \textbf{(C)}, in \texttt{UK-RU}. We only consider users that shared at least 10 tweets. Median values in \textbf{(A)} are 7.9 hours for \emph{Active} and 6.4 hours for \emph{Suspended}. Median values in \textbf{(B)} are 7.9 hours for \emph{Old} and 5.8 hours for \emph{New}. Median values in \textbf{(C)} are 7.9 hours for \emph{Old Active}, 6.5 hours for \emph{Old Suspended}, 6 hours for \emph{New Active} and 4.4 hours for \emph{New Suspended}. In each panel, distributions are statistically different according to two-sided Mann-Whitney tests ($p < 0.001$).}
    \label{fig:tweeting-activity-uk}
\end{figure}

\begin{figure}[!h]
    \centering
    \includegraphics[width=\linewidth]{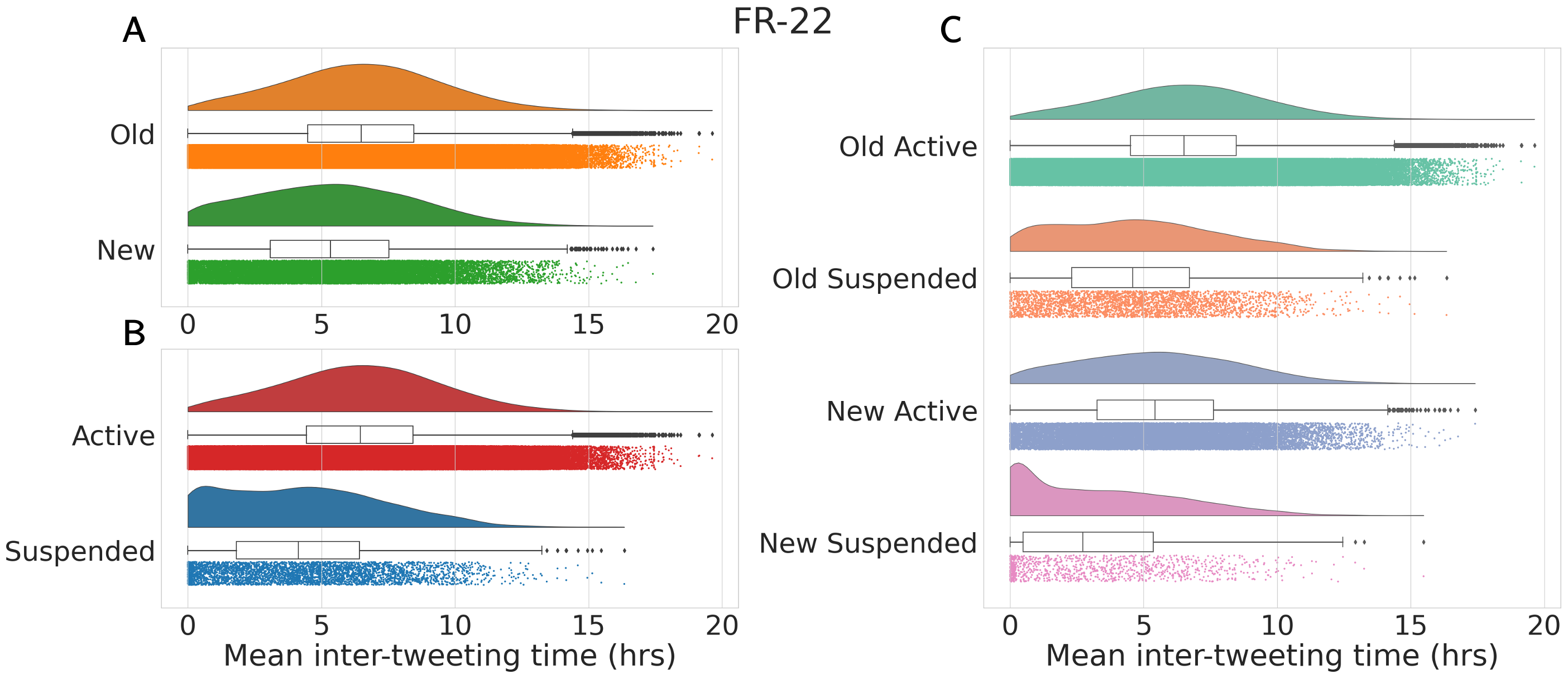}
    \caption{Distribution of the mean inter-tweeting time (in hours) for \emph{Active} versus \emph{Suspended} accounts \textbf{(A)}, \emph{Old} versus \emph{New} users \textbf{(B)} and all four classes \textbf{(C)}, in \texttt{FR-22}. We only consider users that shared at least 10 tweets. Median values in \textbf{(A)} are 6.46 hours for \emph{Active} and 4.13 hours for \emph{Suspended}. Median values in \textbf{(B)} are 6.49 hours for \emph{Old} and 5.33 hours for \emph{New}. Median values in \textbf{(C)} are 6.51 hours for \emph{Old Active}, 4.59 hours for \emph{Old Suspended}, 5.43 hours for \emph{New Active} and 2.72 hours for \emph{New Suspended}. In each panel, distributions are statistically different according to two-sided Mann-Whitney tests ($p < 0.001$).}
    \label{fig:tweeting-activity-fr}
\end{figure}

\begin{figure}[!t]
    \centering
    \includegraphics[width=\linewidth]{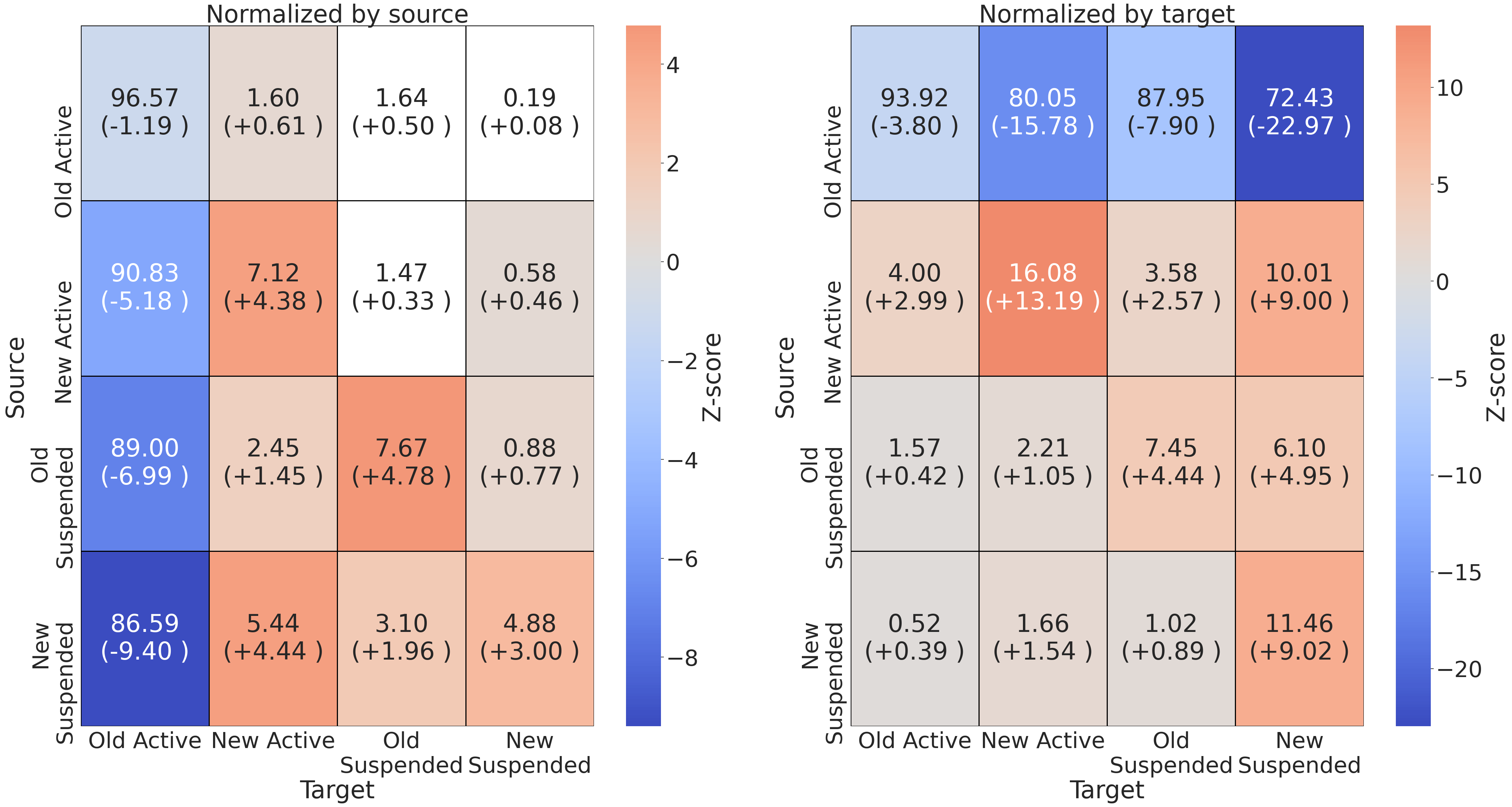}
    \caption{Heatmap of the observed amount of interactions normalized by source \textbf{(left)} and target \textbf{(right)} occurring between different classes of users and, in brackets, the difference with the expected value obtained through the null model, for the \textit{UK-RU} dataset. Colors indicate Z-scores, and we only color cells with Z-score significant at $\alpha=0.05$. Cells are annotated with the normalized volume of interactions, and brackets report the difference w.r.t the value observed in the null model.} 
    \label{fig:interactions-heatmap-uk-ru}
\end{figure}

\begin{figure}[!t]
    \centering
    \includegraphics[width=\linewidth]{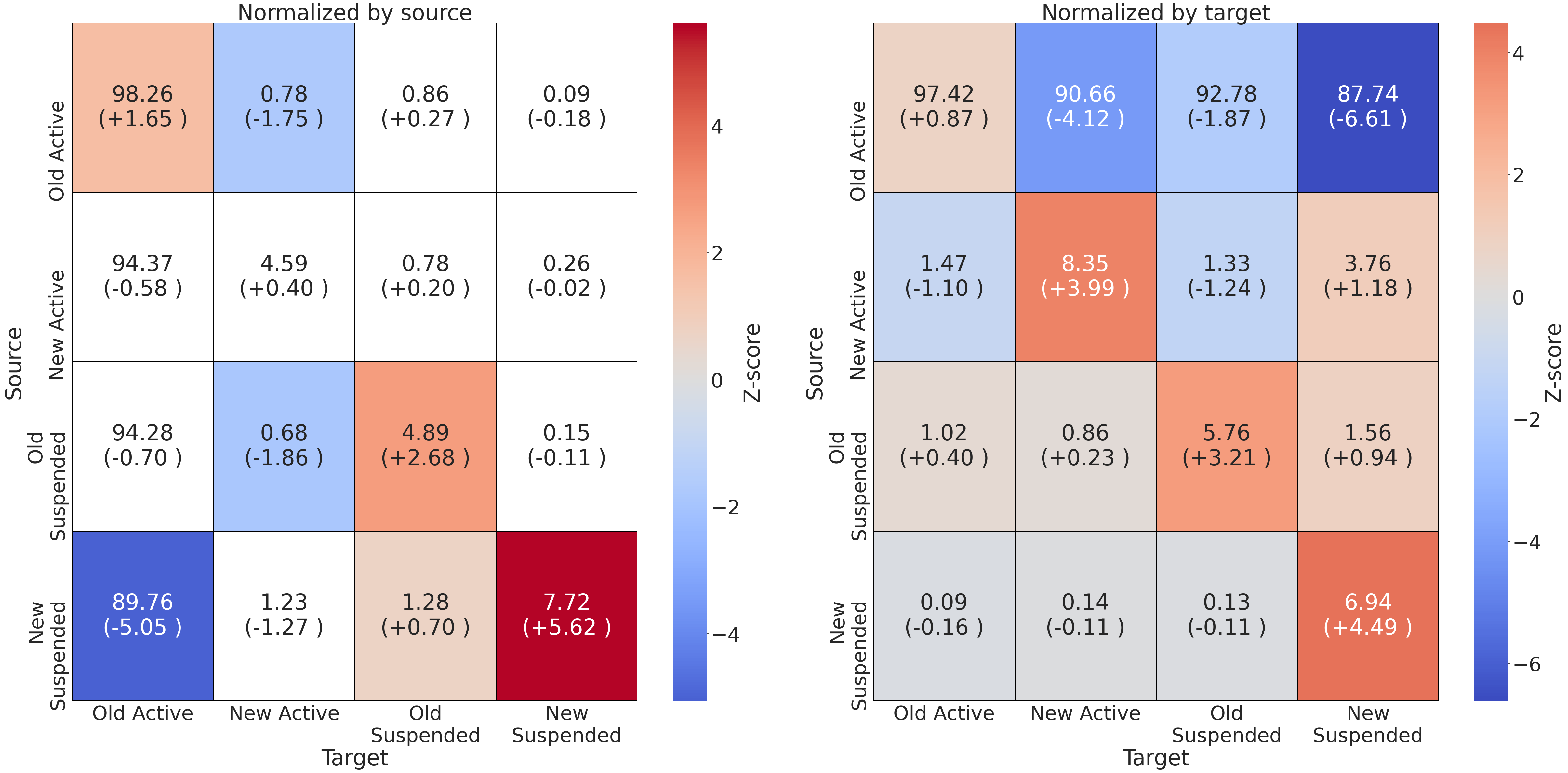}
    \caption{Heatmap of the observed amount of interactions normalized by source \textbf{(left)} and target \textbf{(right)} occurring between different classes of users and, in brackets, the difference with the expected value obtained through the null model, for the \textit{FR-22} dataset. Colors indicate Z-scores, and we only color cells with Z-score significant at $\alpha=0.05$. Cells are annotated with the normalized volume of interactions, and brackets report the difference w.r.t the value observed in the null model.} 
    \label{fig:interactions-heatmap-fr}
\end{figure}

\begin{figure}[!t]
    \centering
    \includegraphics[width=\linewidth]{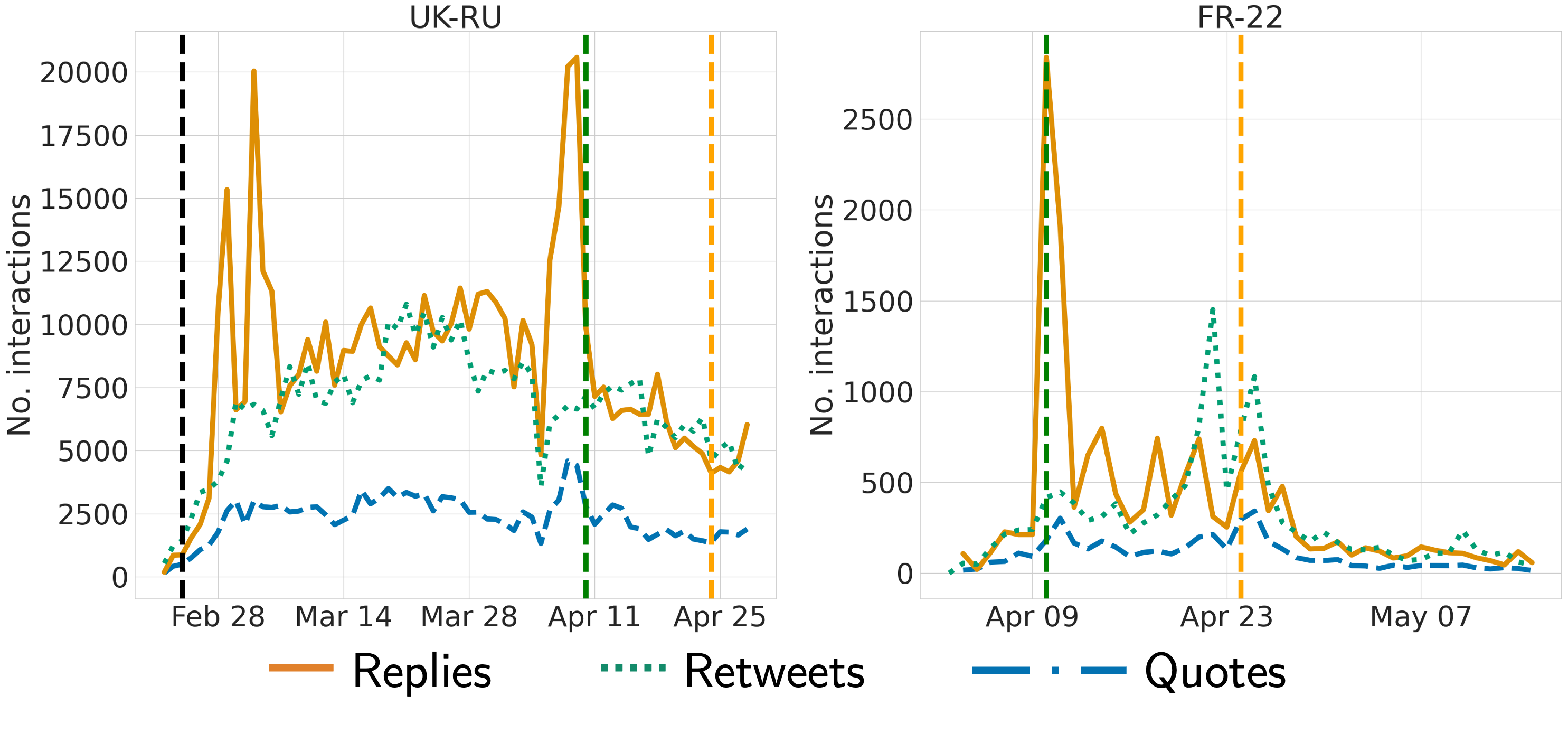}
    \caption{Number of interactions directed from New Suspended accounts to other accounts in \texttt{UK-RU} \textbf{(left)} and \texttt{FR-22} \textbf{(right)}. Vertical lines indicate the invasion of Ukraine (black) and the two rounds of elections (green and light orange)} 
    \label{fig:ts-interactions}
\end{figure}

\begin{figure}[!t]
    \centering
    \includegraphics[width=\linewidth]{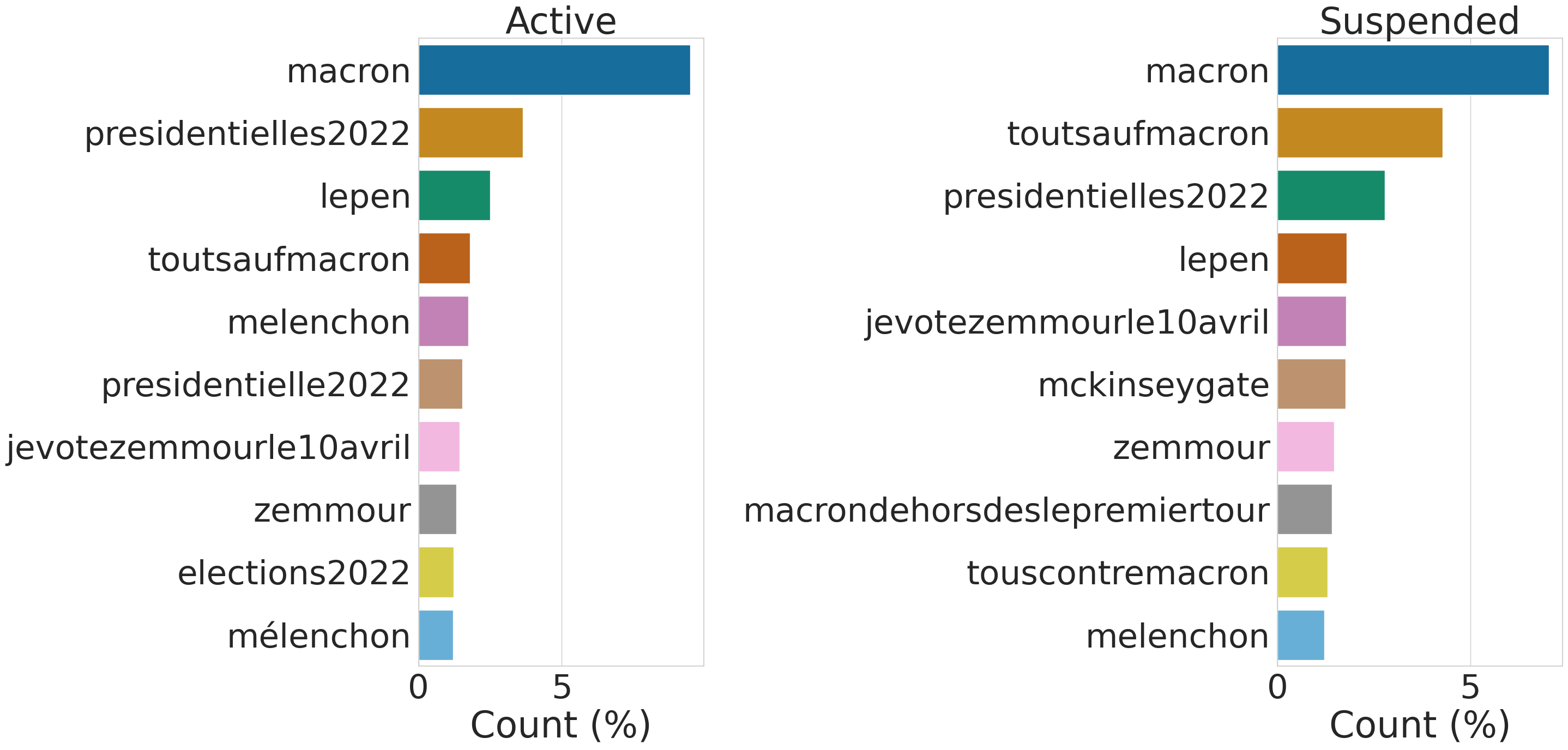}
    \caption{Top-10 most frequent hashtags shared by Suspended and Active users in \texttt{FR-22}.}
    \label{fig:top-hashtags-fr}
\end{figure}

\begin{figure}[!t]
    \centering
    \includegraphics[width=\linewidth]{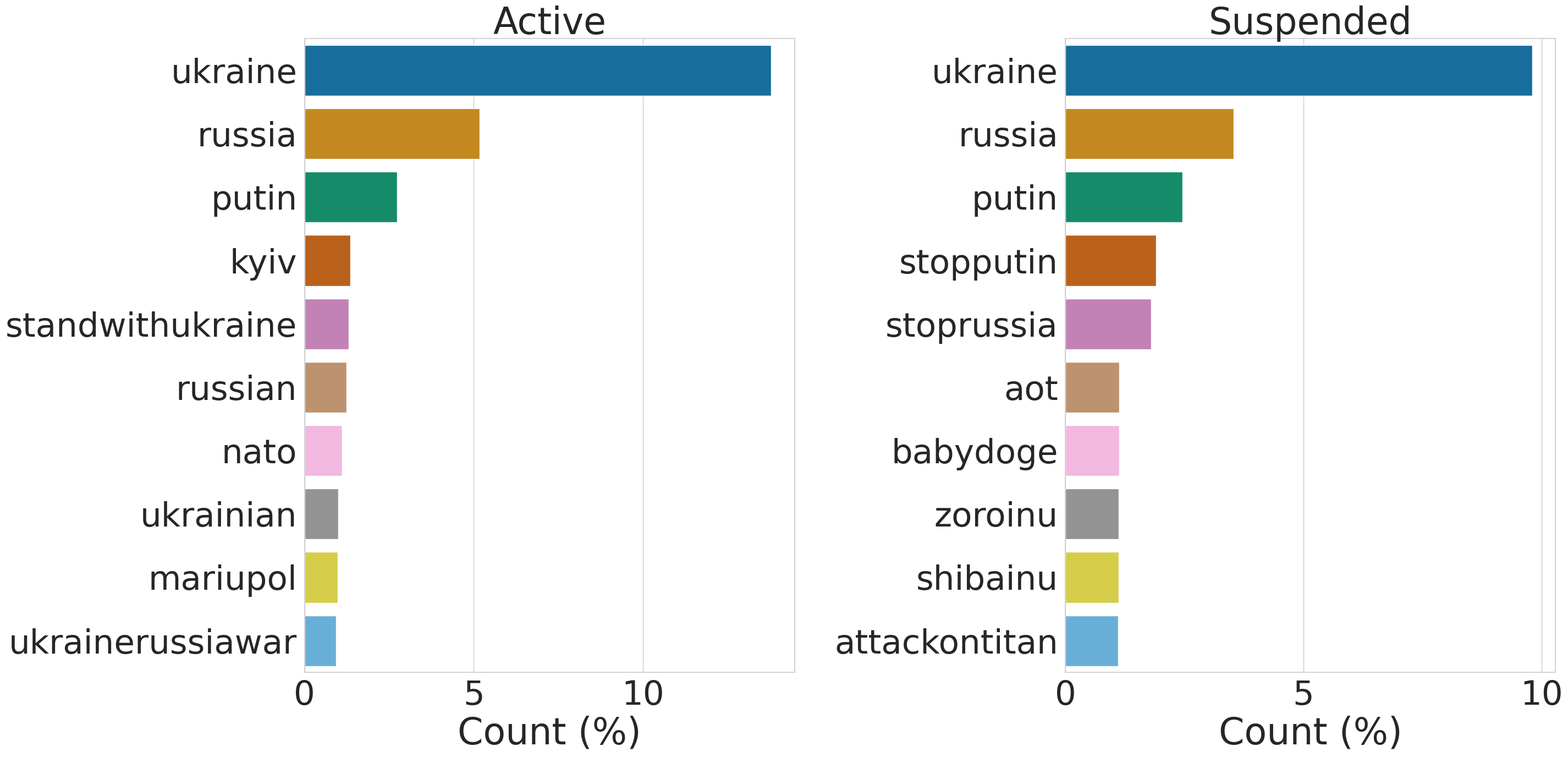}
    \caption{Top-10 most frequent hashtags shared by Suspended and Active users in \texttt{UK-RU}.}
    \label{fig:top-hashtags-uk-ru}
\end{figure}

\begin{figure}[!t]
    \centering
    \includegraphics[width=\linewidth]{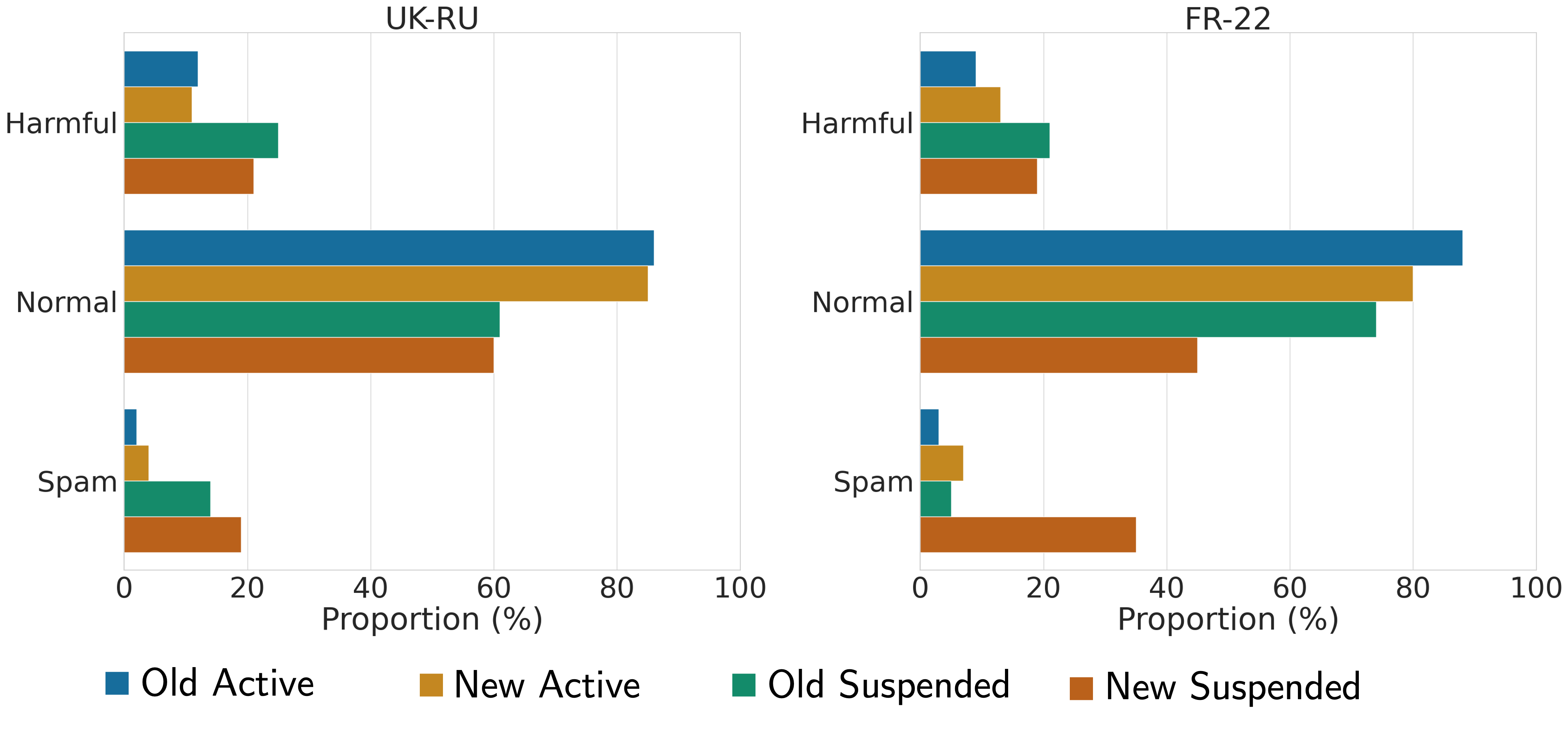}
    \caption{Proportion of tweets that were labeled as Harmful, Normal or Spam for each class of accounts for \textit{UK-RU} \textbf{(left)} and \textit{FR-22} \textbf{(right)}.}
    \label{fig:labeled_tweets}
\end{figure}

%%%%%%%%%%%%%%%%%%%%%%%%%%%%%%%%%%%
%%                               %%
%% Tables                        %%
%%                               %%
%%%%%%%%%%%%%%%%%%%%%%%%%%%%%%%%%%%

%% Use of \listoftables is discouraged.
%%
\section*{Table legends}
\begin{table}
\centering
\begin{tabular}{lllll}
\hline
ukraine & russia & Putin & SlavaUkraini & ukrainian \\
soviet & kremlin & nato & kyev & moscow \\
zelensky & fsb & kgb & donbas & luhansk \\
\hline
\end{tabular}
\caption{Sample of keywords employed to collect tweets relevant to Russia's invasion of Ukraine.}
\label{tab:uk-ru-keywords}
\end{table}

\begin{table}
\centering
\begin{tabular}{lll}
\hline
macron & lepen & zemmour \\
zemmourtrocadero & mckinseygate & mckinseymacrongate  \\
scandalemacron & zemmourbfm & reconquete2022  \\
\hline
\end{tabular}
\caption{Sample of keywords employed to collect tweets relevant to the 2022 French Presidential elections.}
\label{tab:fr-keywords}
\end{table}

\end{backmatter}
\end{document}